\documentclass[a4paper,11pt]{article}
\pdfoutput=1

\usepackage{jheppub}
\usepackage{amsmath,amssymb,amsfonts}
\usepackage{graphicx}
\usepackage{cleveref}
\usepackage{hyperref}
\usepackage{srcltx}
\usepackage{bbm}
\usepackage{mathdots}
\usepackage{color}
\usepackage{dsfont}
\usepackage{slashed}
\usepackage{physics}
\usepackage{subfigure}
\usepackage{tikz}
\usepackage[normalem]{ulem}
\usepackage{ytableau}

\crefname{table}{Table}{Tables}
\crefname{equation}{Eq.}{Eqs.}
\crefname{appendix}{App.}{Apps.}
\crefname{section}{Sec.}{Secs.}
\crefname{figure}{Fig.}{Figs.}

\newcommand{\drawsquare}[2]{\hbox{%
\rule{#2pt}{#1pt}\hskip-#2pt
\rule{#1pt}{#2pt}\hskip-#1pt
\rule[#1pt]{#1pt}{#2pt}}\rule[#1pt]{#2pt}{#2pt}\hskip-#2pt
\rule{#2pt}{#1pt}}

\newcommand{\Yfund}{\raisebox{-.5pt}{\drawsquare{6.5}{0.4}}}
\newcommand{\Ysymm}{\raisebox{-.5pt}{\drawsquare{6.5}{0.4}}\hskip-0.4pt%
        \raisebox{-.5pt}{\drawsquare{6.5}{0.4}}}
\newcommand{\Yasymm}{\raisebox{-3.5pt}{\drawsquare{6.5}{0.4}}\hskip-6.9pt%
        \raisebox{3pt}{\drawsquare{6.5}{0.4}}}
\newcommand{\Ythreea}{\raisebox{-3.5pt}{\drawsquare{6.5}{0.4}}\hskip-6.9pt%
        \raisebox{3pt}{\drawsquare{6.5}{0.4}}\hskip-6.9pt
        \raisebox{9.5pt}{\drawsquare{6.5}{0.4}}}
\newcommand{\Yfoura}{\raisebox{-3.5pt}{\drawsquare{6.5}{0.4}}\hskip-6.9pt%
        \raisebox{3pt}{\drawsquare{6.5}{0.4}}\hskip-6.9pt
        \raisebox{9.5pt}{\drawsquare{6.5}{0.4}}\hskip-6.9pt
        \raisebox{16pt}{\drawsquare{6.5}{0.4}}}
\providecommand{\CC}{\mathcal{C}}
\providecommand{\PP}{\mathcal{P}}

\title{Outer Automorphism Anomalies}

\author[a,b]{Brian Henning,}\emailAdd{brian.henning@epfl.ch}
\author[c]{Xiaochuan Lu,}\emailAdd{xlu@uoregon.edu}
\author[d]{Tom Melia}\emailAdd{tom.melia@ipmu.jp}
\author[d,e,f,1]{and Hitoshi Murayama\note{Hamamatsu Professor}}\emailAdd{hitoshi@berkeley.edu}

\affiliation[a]{Theoretical Particle Physics Laboratory (LPTP), Institute of Physics, EPFL, Lausanne, Switzerland}
\affiliation[b]{D\'epartment de Physique Th\'eorique, Universit\'e de Gen\`{e}ve, 24 quai Ernest-Ansermet, 1211 Gen\`eve 4, Switzerland}
\affiliation[c]{Institute for Fundamental Science, Department of Physics, University of Oregon, Eugene, OR 97403, USA}
\affiliation[d]{Kavli Institute for the Physics and Mathematics of the
  Universe (WPI), University of Tokyo Institutes for Advanced Study, University of Tokyo,
  Kashiwa 277-8583, Japan}
\affiliation[e]{Department of Physics, University of California, Berkeley, CA 94720, USA}
\affiliation[f]{Ernest Orlando Lawrence Berkeley National Laboratory, Berkeley, CA 94720, USA}

\abstract{We discuss anomalies associated with outer automorphisms in gauge theories based on classical groups, namely charge conjugations for $SU(N)$ and parities for $SO(2r)$. We emphasize the inequivalence (yet related by a flavor transformation) between two versions of charge conjugation for $SU(2k)$, $SO(2r)$, and $E_6$ symmetries. The subgroups that commute with the outer automorphisms are identified. Some charge conjugations can lead to a paradox, which is resolved by the observation that they are anomalous and hence not symmetries. We then discuss anomaly matching conditions that involve the charge conjugations or parities. Interesting examples are given where the charge conjugation is spontaneously broken.}

\begin{document}
\maketitle
\flushbottom
\setcounter{page}{2}
\newpage

\section{Introduction}

Charge conjugation symmetry---the interchanging of particles and antiparticles---plays a central role in our understanding of the strong and electromagnetic interactions. It explains, for example, why the neutral pion may decay into two photons, $\pi^0\to2\gamma$, but not three, $\pi^0\not\to3\gamma$. Its violation in the weak interaction is a hallmark of the standard model;  it is at the same time tied to and guides proposed solutions to outstanding questions that the standard model leaves unanswered, including the strong CP problem and the matter-antimatter asymmetry of the universe.

The action of charge conjugation in gauge theories is an example of an outer automorphism of the gauge group; see {\it e.g.}~\cite{Graf:2020yxt}. An outer automorphism of a group $G$ is an automorphism, {\it i.e.} an isomorphism of $G$ onto itself, that can {\it not} be written in the form $g\to h g h^{-1}$, with some fixed $h\in G$. The Lie algebra is mapped onto itself under an automorphism such that the commutation relations are preserved. In general, the outer automorphisms of a Lie algebra can be discerned from the symmetries of the corresponding Dynkin diagram. Charge conjugation is the name reserved for the map that exchanges all representations of the Lie algebra with their complex conjugates. Another example of an outer automorphism is parity which acts by reversing the signs of certain components of the vector representation; see {\it e.g.}~\cite{Henning:2017fpj}.

In this paper we study  subtleties and curiosities of outer automorphisms in gauge theories, with a particular eye to how they behave in the quantum theory, namely, whether or not they are anomalous. Anomalies also play a central role in our understanding of quantum field theory (explaining, for example, the rate of $\pi^0\to2\gamma$ decay~\cite{Adler:1969gk,Bell:1969ts}). The `t Hooft anomaly matching conditions~\cite{tHooft:1979rat} offer a rare and powerful non-perturbative probe of strongly coupled dynamics. Another recent proposal by one of the present authors (HM) uses anomaly-mediated supersymmetry breaking to give a controlled approximation to probe non-supersymmetric strongly coupled systems~\cite{Murayama:2021xfj, Csaki:2021xhi, Csaki:2021aqv, Csaki:2021jax, Csaki:2021xuc}. The study of anomalies for discrete gauge symmetries was pioneered in the works~\cite{IBANEZ1991291, PRESKILL1991207, Banks:1991xj, IBANEZ1993301, Csaki:1997aw, Araki:2006sqx, Araki:2008ek}, with the `t Hooft matching conditions for discrete symmetries studied in~\cite{Csaki:1997aw} by one of the authors (HM); the treatment of outer automorphism anomalies was however missed, with the present work filling this gap. For a modern take on discrete gauge anomalies from the viewpoint of symmetry protected topological phases; see {\it e.g.}~\cite{Hsieh:2018ifc}.

One example of a subtlety we encounter is the not widely known fact that in certain cases there can be two inequivalent versions of charge conjugation. This issue was, to our knowledge, only recently discussed in the literature, in the context of gauging principle extensions of $SU(N)$ gauge theories~\cite{Bourget:2018ond, Arias-Tamargo:2019jyh}. We will see another example of this below in the context of parities for the Spin$(2r)$ symmetries. For the case of charge conjugation in $SU(N)$, for even $N$ we can define both a symmetric and an anti-symmetric version of charge conjugation.  This gives rise to an apparent paradox in QCD, since the anti-symmetric charge conjugation symmetry should forbid the expected chiral condensate that dynamically breaks chiral symmetry. We will see  the resolution to this paradox is that the anti-symmetric version of charge conjugation symmetry is anomalous. The two versions are shown explicitly to be  related by a flavor transformation which explains the difference in anomalies.

For the cases where the outer automorphism is non-anomalous under the gauge group, studying the anomalies associated with it leads to a variety of interesting and important conclusions and consistency conditions. We will illustrate this in numerous examples. We study the anomaly matching conditions that must hold for well-established dualities in $\nobreak{\mathcal{N}=1}$ supersymmetric QCD-like gauge theories. Specifically we consider charge conjugation anomalies in Seiberg duality~\cite{Seiberg:1994pq}, Kutasov duality~\cite{Kutasov:1995ve}, and a duality due to Intriligator, Leigh, and Strassler~\cite{Intriligator:1995ax}; we also consider parity outer automorphism anomalies in a duality studied by Intriligator and Seiberg~\cite{Intriligator:1995id}. In all cases we observe the matching conditions to hold. We also check an example of an $\mathcal{N}=1$ supersymmetric s-confining theory~\cite{Csaki:1996zb} with a parity outer automorphism, and confirm that the anomalies between UV and IR match, as by definition they must.

We further consider two examples  of confining supersymmetric theories to study the spontaneous breaking of outer automorphisms. On the face of it, the two theories---an $SO(6)$ theory (from~\cite{Intriligator:1995id}) that breaks to $SU(2)\times SU(2)$, and an $SU(6)$ theory (from~\cite{Csaki:1996zb}) that breaks to $SU(3)\times SU(3)$---have similar dynamics. However, we find that only in the case of the $SO(6)$ theory, the outer automorphism anomalies do not match between UV and IR, indicating its spontaneous breakdown in this case. It leads to this theory having two ground states, and to the possibility of domain walls.

The paper has the following outline. We first give an overview of the outer automorphisms of simple Lie algebras in \cref{sec:general}, with an emphasis on  three different types of equivalence relations among them. In \cref{sec:twov}, we provide a concrete discussion of the two inequivalent versions of the charge conjugation for $SU(N)$ symmetries. \cref{sec:parity} is dedicated to a similar discussion on the parities for general $SO(2r)$ symmetries. In \cref{sec:E6}, we briefly discuss the case of $E_6$ and the special triality of $SO(8)$. In \cref{sec:para}, we discuss a paradox that occurs in QCD-like $SU(N)$ theories for one of the definitions of charge conjugation, and provide its resolution. In \cref{sec:anom}, we first review the results of~\cite{Csaki:1997aw} on discrete anomaly matching, and then study the anomaly matching conditions for outer automorphism symmetries in supersymmetric QCD-like gauge theories (SQCD), finding they are satisfied in all cases. In \cref{sec:spont}, we study the possibility of spontaneous breaking of charge conjugation in two different $\mathcal{N}=1$ supersymmetric gauge theories, finding that in one case spontaneous breaking does occur. We conclude in \cref{sec:conc} with a discussion on the consequence of our results for gauge theories of the principle extensions of $SU(N)$ by charge conjugation, and possible connections to topological protected states.

\section{Outer Automorphisms}
\label{sec:general}

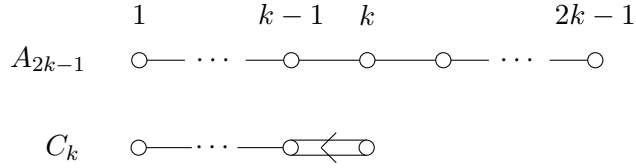
\begin{figure}[t]
\centering
\subfigure{
\begin{tikzpicture}
	\draw (0,0) -- (0.6,0);
	\draw (1.4, 0) -- (4.6,0);
	\draw (5.4, 0) -- (6,0);	
	\draw[fill=white] (0,0) circle(.1);
	\draw[fill=white] (2,0) circle(.1);
	\draw[fill=white] (3,0) circle(.1);
	\draw[fill=white] (4,0) circle(.1);
	\draw[fill=white] (6,0) circle(.1);	
	\node at (1,0) {$\cdots$};
	\node at (5,0) {$\cdots$};
	\node at (0,0.6) {$1$};	
	\node at (2,0.6) {$k-1$};	
	\node at (3,0.6) {$k$};	
	\node at (6,0.6) {$2k-1$};	
	\node at (-1.2,0) {$A_{2k-1}$};	
\end{tikzpicture}
}\\[10pt]
\subfigure{
\begin{tikzpicture}
	\draw (0,0) -- (0.7,0);
	\draw (1.3, 0) -- (2,0);
	\draw (2,0.1) -- (3,0.1);
	\draw (2,-0.1) -- (3,-0.1);
	\draw (2.6,0.2) -- (2.4,0);
	\draw (2.6,-0.2) -- (2.4,0);		
	\draw[fill=white] (0,0) circle(.1);
	\draw[fill=white] (2,0) circle(.1);
	\draw[fill=white] (3,0) circle(.1);
	\draw[fill=white] (6.2,0) circle(.0);	
	\node at (1,0) {$\cdots$};
	\node at (-1,0) {$C_k$};	
\end{tikzpicture}
}
\caption{Dynkin diagram $A_{2k-1}$ (Lie algebra $\frak{su}(N=2k)$) has a $\mathbb{Z}_2$ outer automorphism, which flips the order of its nodes. Folding it by \emph{average} yields the Dynkin diagram $C_k$, which represents its $\CC_A$-invariant subalgebra $\frak{sp}(2k)$.}
\label{fig:A2km1folding}
\end{figure}

\begin{figure}[t]
\centering
\subfigure{
\begin{tikzpicture}
	\draw (0,0) -- (1,0);
	\draw (1,0) -- (2,-.5);
	\draw (1,0) -- (2,.5);
	\draw[fill=white] (0,0) circle(.1);
	\draw[fill=white] (1,0) circle(.1);
	\draw[fill=white] (2,-.5) circle(.1);
	\draw[fill=white] (2,.5) circle(.1);
	\node at (-1,0) {$D_4$};
\end{tikzpicture}
}\\[10pt]
\subfigure{
\begin{tikzpicture}
    \draw (0,0) -- (1,0);
	\draw (0,0.1) -- (1,0.1);
	\draw (0,-0.1) -- (1,-0.1);
	\draw (0.6,0.2) -- (0.4,0);
	\draw (0.6,-0.2) -- (0.4,0);		
	\draw[fill=white] (0,0) circle(.1);
	\draw[fill=white] (1,0) circle(.1);
	\node at (-1,0) {$G_2$};
\end{tikzpicture}
}
\caption{Dynkin diagram $D_4$ (Lie algebra $\frak{so}(8)$) has a triality $S_3$ outer automorphism which permutates its three branches. Folding it by \emph{average} yields the Dynkin diagram $G_2$.}
\label{fig:D4folding}
\end{figure}
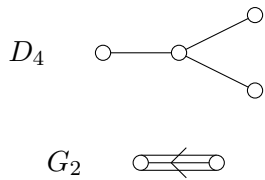

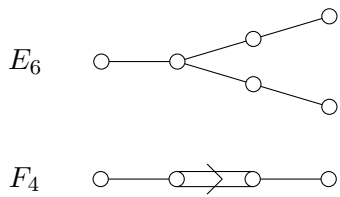
\begin{figure}[t]
\centering
\subfigure{
\begin{tikzpicture}
	\draw (0,0) -- (1,0);
	\draw (1,0) -- (3,-.6);
	\draw (1,0) -- (3,.6);
	\draw[fill=white] (0,0) circle(.1);
	\draw[fill=white] (1,0) circle(.1);
	\draw[fill=white] (2,-.3) circle(.1);
	\draw[fill=white] (2,.3) circle(.1);
	\draw[fill=white] (3,-.6) circle(.1);
	\draw[fill=white] (3,.6) circle(.1);
	\node at (-1,0) {$E_6$};
\end{tikzpicture}
}\\[10pt]
\subfigure{
\begin{tikzpicture}
	\draw (0,0) -- (1,0);
	\draw (2, 0) -- (3,0);
	\draw (1,0.1) -- (2,0.1);
	\draw (1,-0.1) -- (2,-0.1);
	\draw (1.4,0.2) -- (1.6,0);
	\draw (1.4,-0.2) -- (1.6,0);		
	\draw[fill=white] (0,0) circle(.1);
	\draw[fill=white] (1,0) circle(.1);
	\draw[fill=white] (2,0) circle(.1);
	\draw[fill=white] (3,0) circle(.1);
	\node at (-1,0) {$F_4$};
\end{tikzpicture}
}
\caption{The Dynkin diagram $E_6$ has a $\mathbb{Z}_2$ outer automorphism which interchanges its two branches. Folding it by \emph{average} yields its $\CC$-invariant subalgebra $F_4$.}
\label{fig:E6folding}
\end{figure}

It is well known that outer automorphisms of simple Lie algebras correspond to the symmetries of the corresponding Dynkin diagrams. Therefore, they exist only for
\begin{equation*}\renewcommand\arraystretch{1.3}
\begin{array}{ll}
SU(N)  &\qquad \text{\cref{fig:A2km1folding,fig:A2kfolding}} \\
SO(2r) &\qquad \text{\cref{fig:Drfolding,fig:D4folding}} \\
E_6    &\qquad \text{\cref{fig:E6folding}}
\end{array}
\end{equation*}
For all these groups but $SO(8)$, the outer automorphism is a ${\mathbb Z}_{2}$ group, which we will call
\begin{equation*}
\text{``charge conjugation'':}\;\;\;   \mathbb{Z}_2 = \{1\,,\,\CC\}
\qquad\text{or}\qquad
\text{``parity'':}\;\;\;               \mathbb{Z}_2 = \{1\,,\,\PP\}
\end{equation*}
interchangeably. We also accept the abuse of these terms to refer to both the group $\mathbb{Z}_2 = \{1\,,\,\CC\}$ ($\mathbb{Z}_2 = \{1\,,\,\PP\}$) and the element $\CC$ ($\PP$). For $SO(8)$, the outer automorphism group is $S_{3}$ which is called triality. Yet for our applications we will be only interested in the ${\mathbb Z}_{2}$ subgroup of $S_{3}$ and use the same terminology. Unfortunately, we could not find mathematical literature that fleshed out how the outer automorphisms act explicitly on each Lie algebra. So we briefly describe it in this section.

\begin{figure}
\centering
\subfigure{
\begin{tikzpicture}
	\draw (0,0) -- (0.6,0);
	\draw (1.4, 0) -- (5.6,0);
	\draw (6.4, 0) -- (7,0);	
	\draw[fill=white] (0,0) circle(.1);
	\draw[fill=white] (2,0) circle(.1);
	\draw[fill=white] (3,0) circle(.1);
	\draw[fill=white] (4,0) circle(.1);
	\draw[fill=white] (5,0) circle(.1);	
	\draw[fill=white] (7,0) circle(.1);		
	\node at (1,0) {$\cdots$};
	\node at (6,0) {$\cdots$};
	\node at (0,0.6) {$1$};		
	\node at (3,0.6) {$k$};	
	\node at (4,0.6) {$k+1$};	
	\node at (7,0.6) {$2k$};	
	\node at (-1,0) {$A_{2k}$};
\end{tikzpicture}
}\\[10pt]
\subfigure{
\begin{tikzpicture}
	\draw (0,0) -- (0.7,0);
	\draw (1.3, 0) -- (2,0);
	\draw (2,0.1) -- (3,0.1);
	\draw (2,-0.1) -- (3,-0.1);
	\draw (2.6,0) -- (2.4,0.2);
	\draw (2.6,0) -- (2.4,-0.2);		
	\draw[fill=white] (0,0) circle(.1);
	\draw[fill=white] (2,0) circle(.1);
	\draw[fill=white] (3,0) circle(.1);	
	\draw[fill=white] (7.25,0) circle(.0);	
	\node at (1,0) {$\cdots$};
	\node at (-1,0) {$B_k$};
\end{tikzpicture}
}
\caption{Dynkin diagram $A_{2k}$ (Lie algebra $\frak{su}(N=2k+1)$) has a $\mathbb{Z}_2$ outer automorphism, which flips the order of its nodes. Folding it by \emph{average} yields the Dynkin diagram $B_k$, which represents its $\CC_S$-invariant subalgebra $\frak{so}(2k+1)$.}
\label{fig:A2kfolding}
\end{figure}

\begin{figure}[t]
\centering
\subfigure{
\begin{tikzpicture}\label{fig:ff}
	\draw (0,0) -- (2.7,0);
	\draw (3.3, 0) -- (4,0);
	\draw (4,0) -- (5,-.5);
	\draw (4,0) -- (5,.5);
	\draw[fill=white] (0,0) circle(.1);
	\draw[fill=white] (1,0) circle(.1);
	\draw[fill=white] (2,0) circle(.1);
	\draw[fill=white] (4,0) circle(.1);
	\draw[fill=white] (5,-.5) circle(.1);
	\draw[fill=white] (5,.5) circle(.1);
	\node at (3,0) {$\cdots$};
	\node at (-1,0) {$D_r$};
\end{tikzpicture}
}\\[10pt]
\subfigure{
\begin{tikzpicture}
	\draw (0,0) -- (2.7,0);
	\draw (3.3, 0) -- (4,0);
	\draw (4,0.1) -- (5,0.1);
	\draw (4,-0.1) -- (5,-0.1);
	\draw (4.4,0.2) -- (4.6,0);
	\draw (4.4,-0.2) -- (4.6,0);		
	\draw[fill=white] (0,0) circle(.1);
	\draw[fill=white] (1,0) circle(.1);
	\draw[fill=white] (2,0) circle(.1);
	\draw[fill=white] (4,0) circle(.1);
	\draw[fill=white] (5,0) circle(.1);	
	\node at (3,0) {$\cdots$};
	\node at (-1,0) {$B_{r-1}$};
\end{tikzpicture}
}
\caption{Dynkin diagram $D_r$ (Lie algebra $\frak{so}(2r)$) has a $\mathbb{Z}_2$ outer automorphism which interchanges its two branches. Folding it by \emph{average} yields the Dynkin diagram $B_{r-1}$, which represents its $\PP$-invariant subalgebra $\frak{so}(2r-1)$.}
\label{fig:Drfolding}
\end{figure}
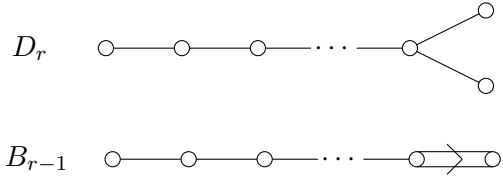

On the Lie algebra, an outer automorphism leaves a subalgebra invariant while all the other elements are odd. Namely it is an involution of the Lie algebra. We will refer to the subalgebra that is left invariant as the ``$\CC$-invariant'' or ``$\PP$-invariant'' subalgebra (subgroup). For some involutions, the $\CC$-invariant subalgebra can be obtained by folding the Dynkin diagram, as shown in \cref{fig:A2km1folding,fig:A2kfolding,fig:Drfolding,fig:D4folding,fig:E6folding}.\footnote{See \cite{saito1985extended, Fuchs:1996vp, Fuchs:1996ju} as well as App.~B in \cite{Graf:2020yxt} and App.~C.2 in \cite{Henning:2017fpj} for details of folding, in particular, the two types of folding: folding by \emph{average} and folding by \emph{sum}.} Lie algebra involutions have been systematically classified as symmetric spaces $G/K$ (\textit{e.g.} \cite{Arias-Tamargo:2019jyh}), but in general only a small subset of symmetric spaces may give outer automorphisms on the group $G$. In addition, the outer automorphism on $G$ must interchange certain representations indicated by the symmetry of the Dynkin diagram, which become equivalent under $K$:
\begin{equation*}\renewcommand\arraystretch{1.3}
\begin{array}{ll}
SU(N)  &\qquad \text{fundamental and anti-fundamental representations} \\
SO(2r) &\qquad \text{two inequivalent spinor representations} \\
E_6    &\qquad \text{${\bf 27}$ and ${\bf 27}^{*}$ representations}
\end{array}
\end{equation*}
These requirements leave us with only the following possibilities for $G/K$ to form an outer automorphism:
\begin{equation}\renewcommand\arraystretch{1.3} \label{eq:possibilities}
\begin{array}{l}
	SU(N)/SO(N) \\
	SU(2k)/Sp(2k) \\
	SO(2r)/\left( SO(q)\times SO(2r-q) \right) \qquad (q~\mbox{odd}) \\
	E_{6}/(Sp(8)/{\mathbb Z}_{2}) \\
	E_{6}/F_{4}
\end{array}
\end{equation}

Before discussing each of these outer automorphisms at length, let us summarize some general properties of them. We find that there are three different types of ``equivalence relations'' that can be discussed regarding two outer automorphisms $\CC_1$ and $\CC_2$ of the same group $G$:
\begin{enumerate}
  \item They are equivalent representatives of the quotient group $\text{Aut}(G)/\text{Inn}(G)$, namely that they yield the same coset $\text{Inn}(G)\circ\CC_1=\text{Inn}(G)\circ\CC_2$. Operationally, this condition is the same as requiring $\exists\, g\in G$ such that $\left(g\,\CC_1\right) g_0 \left(g\,\CC_1\right)^{-1} = \CC_2\,g_0\,\CC_2^{-1}$, $\forall g_0 \in G$.
  \item They yield equivalent semidirect product (extension) groups, \textit{i.e.} $G\rtimes \{1\,,\,\CC_1\} = G\rtimes \{1\,,\,\CC_2\}$ in the sense that the two extension groups are isomorphic. Operationally, this condition is the same as requiring $\exists\, g\in G$ such that $g\,\CC_1=\CC_2$.
  \item They are equivalent upon basis change (\textit{i.e.} $G$-conjugation). Operationally, this condition is the same as requiring $\exists\, g\in G$ such that $g\,\CC_1\,g^{-1}=\CC_2$.
\end{enumerate}
The above conditions become stronger in order, \textit{i.e.} later ones are sufficient to guarantee the former ones\footnote{$3\Rightarrow 2$ because $\CC_2 = g\,\CC_1\,g^{-1} = \left( g\,\CC_1\,g^{-1}\,\CC_1 \right) \CC_1 = g'\,\CC_1$.}
\begin{equation}
3 \;\;\Rightarrow\;\;   2 \;\;\Rightarrow\;\;   1 \,,
\end{equation}
but not the other way around. Therefore, two outer automorphisms that are equivalent under a former criteria, may still be inequivalent under a latter criterion. In fact, all the outer automorphisms for the same group $G$ are equivalent under criterion 1, and they are all represented by the symmetry of the Dynkin diagram of $G$. However, under criterion 2, they may fall into inequivalent categories. As we will see, in the case of $SU(2k)$, we indeed find two versions of charge conjugations, $\CC_S$ and $\CC_A$, that are inequivalent under criterion 2. A similar story holds for the case of parities for $SO(2r)$. Under criterion 3, there could be many more inequivalent charge conjugations or parities. A few remarks about the implications of each of these equivalence criteria:
\begin{itemize}
\item Outer automorphisms that are equivalent under criterion 3 are obviously equivalent physical observables and will lead to the same physical consequences. Since their difference completely stems from different basis choices, they are not distinguishable.
\item Outer automorphisms that are not equivalent under criterion 3 are not the same physical observable. In this case, they may correspond to different involutions of the Lie algebra, and their $\CC$-invariant ($\PP$-invariant) subalgebras may be different. Nevertheless, if they are still equivalent under criterion 2, then they give the same extension group of $G$. In this case, we expect them to have the same anomaly properties and lead to the same anomaly matching condition.
\item Outer automorphisms that are not equivalent under criterion 2 can in principle have different anomaly properties and yield different anomaly matching conditions. As we will see, this indeed happens for the charge conjugations $\CC_S$ and $\CC_A$.
\item We will see that of the possible symmetric spaces $SO(2r)/\left( SO(q)\times SO(2r-q) \right)$ (for $q$ odd) in \cref{eq:possibilities},  only two correspond to inequivalent outer automorphims under criterion 2.
\end{itemize}

\section{Two Versions of Charge Conjugation for \boldmath$SU(N)$}
\label{sec:twov}

How do we extend an $SU(N)$ symmetry with an outer automorphism which we call charge conjugation? Here, the charge conjugation $\CC$ is meant to be the outer automorphism of the $\mathfrak{su}(n)$ Lie algebra which interchanges a representation with its complex conjugate representation.  The symmetry group is then the semi-direct product $SU(N) \rtimes \CC$; see {\it e.g.}~for its application in QCD chiral Lagrangian. Here we demonstrate that there is a unique definition of charge conjugation for $SU(N)$ with odd $N$ up to basis changes, but there are two inequivalent definitions of charge conjugation for $SU(N)$ with even $N$.  This point is consistent with the papers \cite{Bourget:2018ond, Arias-Tamargo:2019jyh} where they discussed gauging the {\it principal extension}\/ $SU(N) \rtimes \CC$.  Here we present the discussion which is very concrete compared to previous literature.\footnote{The anomalies associated with charge conjugation was incorrectly dismissed in~\cite{Csaki:1997aw}} Our result confirms the argument in the previous section \ref{sec:general} that the $\CC$-invariant subgroup under charge conjugation is either $SO(N)$ or $Sp(N)$ if $N$ is even.

\subsection{Requirements}

The charge conjugation is an operation that interchanges the fundamental representation and the anti-fundamental representation which has the following properties:
\begin{enumerate}
\item linear
\item unitary
\item $\CC^{2} = 1$
\item compatible with $SU(N)$
\end{enumerate}
We will see below concrete realizations of these requirements.

\subsection{Fundamental and Anti-Fundamental Representations}

We start with the fundamental representation ${\bf N}$ of $SU(N)$. It is a complex representation and the charge conjugation is not closed within this representation. The charge conjugation works on the direct sum ${\bf N} \oplus \overline{\bf N}$, and because it interchanges ${\bf N}$ and $\overline{\bf N}$ and is linear, it can be written as
\begin{align}
\CC \left( \begin{array}{c} {\bf N} \\ \overline{\bf N} \end{array} \right)
= \left( \begin{array}{cc} 0 & C_{-}\\ C_{+} & 0 \end{array} \right)
\left( \begin{array}{c} {\bf N} \\ \overline{\bf N} \end{array} \right) \,.
\end{align}
Here, $C_{\pm}$ are $N$ by $N$ matrices. Strictly speaking, the matrix $\CC$ here is a representation matrix of the abstract operation $\CC$ on ${\bf N} \oplus \overline{\bf N}$, but we accept the abuse of notation. Once the matrix $\CC$ is specified on ${\bf N} \oplus \overline{\bf N}$, it can be generalized to any other representations of $SU(N)$ because they are all obtained by tensor products of ${\bf N}$ and $\overline{\bf N}$.

The unitarity requirement is
\begin{align}
\CC^{\dagger} \CC =
\left( \begin{array}{cc} 0 & C_{+}^{\dagger}\\ C_{-}^{\dagger} & 0 \end{array} \right)
\left( \begin{array}{cc} 0 & C_{-}\\ C_{+} & 0 \end{array} \right) = 1 \,.
\end{align}
Namely,
\begin{align}
C_{-}^{\dagger} C_{-} = C_{+}^{\dagger} C_{+} = 1 \,,
\end{align}
and hence $C_{\pm}$ are unitarity.  On the other hand, it also needs to square to unity,
\begin{align}
\CC^{2} =
\left( \begin{array}{cc} 0 & C_{-}\\ C_{+} & 0 \end{array} \right)
\left( \begin{array}{cc} 0 & C_{-}\\ C_{+} & 0 \end{array} \right) = 1 \,,
\end{align}
and hence
\begin{align}
C_{-} C_{+} = C_{+} C_{-} = 1 \,,
\end{align}
or
\begin{align}
C_{-} = C_{+}^{-1} = C_{+}^{\dagger} \,.
\end{align}
We henceforth use the notation $C_+=C$ and $C_- = C^\dagger$. We will also use the tensor notation where the fundamental representation comes with an upper index, and the anti-fundamental representation with a lower index. Then the matrix $C$ has indices $C_{ij}$ while the inverse matrix $(C^{\dagger})^{ij}$.

Now we discuss the compatibility with $SU(N)$, which means for any $g \in SU(N)$, $\CC g \CC$ must also be an element of $SU(N)$. Focusing on unitary representations, any element of $g \in SU(N)$ can be represented on ${\bf N} \oplus \overline{\bf N}$ as
\begin{align}
U(g) \left( \begin{array}{c} {\bf N} \\ \overline{\bf N} \end{array} \right)
= \left( \begin{array}{cc} e^{i T^{a} \omega^{a}} & 0 \\
	0 & V^{\dagger} e^{-i T^{aT} \omega^{a}} V \end{array} \right)
\left( \begin{array}{c} {\bf N} \\ \overline{\bf N} \end{array} \right) \,.
\label{eq:U(g)}
\end{align}
Here $T^{a}$ are traceless hermitian matrices forming the fundamental representation of the $\mathfrak{su}(N)$ Lie algebra and $\omega^{a}$ are real parameters. The anti-fundamental representation should be equivalent to $\left(e^{i T^{a} \omega^{a}}\right)^* = e^{-i T^{aT} \omega^{a}}$ up to a unitary transformation $V$,\footnote{
Note that if the matrices $T^{a}$ form a representation of the Lie algebra $[T^{a}, T^{b}] = i f^{abc} T^{c}$, so will their negative transposed matrices $-T^{aT}$: $[- T^{aT}, -T^{bT}] = i f^{abc} (-T^{cT})$.}
which can be set to $1$ without loss of generality by further changing the basis for the anti-fundamental representation. Then the element $\CC g \CC$ is represented as
\begin{align}
\CC U(g) \CC
&= \left( \begin{array}{cc} 0 & C^{\dagger}\\ C & 0 \end{array} \right)
\left( \begin{array}{cc} e^{i T^{a} \omega^{a}} & 0 \\
	0 & e^{-i T^{aT} \omega^{a}} \end{array} \right)
\left( \begin{array}{cc} 0 & C^{\dagger}\\ C & 0 \end{array} \right)
\notag\\[3pt]
&= \left( \begin{array}{cc}
	C^{\dagger} e^{-i T^{aT} \omega^{a}} C & 0 \\
	0 & C e^{i T^{a} \omega^{a}} C^{\dagger}
	\end{array} \right) \,.
\label{eq:CgC}
\end{align}
For this to be an element of $SU(N)$, there must be a parameter set $\eta^{a}$ such that
\begin{align}
\left( \begin{array}{cc}
	C^{\dagger} e^{-i T^{aT} \omega^{a}} C & 0 \\
	0 & C e^{i T^{a} \omega^{a}} C^{\dagger}
	\end{array} \right)
= \left( \begin{array}{cc}
	e^{i T^{a} \eta^{a}} & 0 \\
	0 & e^{-i T^{aT} \eta^{a}}
	\end{array} \right) \,.
\label{eq:compatibility}
\end{align}
From the upper block in \cref{eq:compatibility}, we find
\begin{align}
e^{-i T^{aT} \omega^{a}} & = C e^{i T^{a} \eta^{a}} C^{\dagger} \,,
\end{align}
and its complex conjugate is
\begin{align}
e^{i T^{a} \omega^{a}} &= C^{*} e^{-i T^{aT} \eta^{a}} C^{T} \,.
\end{align}
Inserting the right-hand side of the lower block in \cref{eq:compatibility} into the right-hand side above, we obtain
\begin{align}
e^{i T^{a} \omega^{a}} &= C^{*} C e^{i T^{a} \omega^{a}} C^{\dagger} C^{T} \,.
\end{align}
By multiplying $C^{*} C$ from the right on both sides of the equation, we see that
\begin{align}
e^{i T^{a} \omega^{a}} C^{*} C &= C^{*} C e^{i T^{a} \omega^{a}} \,.
\end{align}
Namely $C^{*} C$ commutes with {\it any}\/ element $e^{i T^{a} \omega^{a}} \in SU(N)$, while they form the irreducible (fundamental) representation.  Therefore, Schur's lemma says $C^{*} C$ must be proportional to an identity matrix
\begin{align}
C^{*} C = a\,\mathds{1} \,,
\end{align}
and hence
\begin{align}
C = a\, C^{T}   \quad\Rightarrow\quad   C^T = a\, C \,.
\end{align}
Subbing the second expression into the first, we get $C = a^{2}C$ which dictates $a=\pm 1$. On the other hand,
\begin{align}
\det C = a^{N} \det C^{T} = a^{N} \det C \,,
\end{align}
and hence $a^{N} = 1$.  For $N$ odd, the only possibility is $a=1$.  For $N$ even, we have two possibilities $a=\pm 1$.

\subsection{Symmetric or Anti-symmetric Charge Conjugations}

We found $C$ has to be symmetric for $N$ odd, and either symmetric or anti-symmetric for $N$ even.  It is important to note that the symmetry property is independent of the choice of the basis.

When we change the basis of $N \rightarrow U^{\dagger} N$ and $N^{*} \rightarrow U^{T} N^{*}$ representations, the $\CC$ matrix changes as
\begin{align}
	\CC' =
	\left( \begin{array}{cc} U & 0 \\ 0 & U^{*} \end{array} \right)
	\left( \begin{array}{cc} 0 & C^{\dagger} \\ C & 0 \end{array} \right)
	\left( \begin{array}{cc} U^{\dagger} & 0 \\ 0 & U^{T} \end{array} \right)
	= \left( \begin{array}{cc} 0 & U C^{\dagger} U^{T} \\
	U^{*} C U^{\dagger} & 0 \end{array} \right) \,.
\end{align}
We can identify the matrix $C'$ in the new basis
\begin{align}
C' = U^* C U^\dagger \,.
\label{eqn:Ctransform}
\end{align}
The fact that this is {\it not}\/ the usual unitary transformation is important.  When $C$ is symmetric $C^T = C$ in a basis, it is also symmetric in any other bases.  The same is true when $C$ is anti-symmetric $C^T = -C$.  Whether $C$ is symmetric or anti-symmetric is hence basis-independent and makes a difference. Therefore, $SU(N)$ groups admit two inequivalent definitions of charge conjugation when $N$ is even.  We call them $\CC_{S}$ for $C_{S}^{T}=C_{S}$ and $\CC_{A}$ for $C_{A}^{T}=-C_{A}$.

With a \emph{special} unitary transformation $U$, any complex symmetric matrix can be brought to the real positive-semi-definite diagonal matrix, up to a phase factor:
\begin{align}
C_S = U^T e^{i\theta_S} \left( \begin{array}{cccc}
	c_{1}& & &\\ & c_{2} & & \\ & & \ddots & \\ & & & c_{N} \end{array} \right) U \,,
\end{align}
with $c_i\ge0$. Given that $C$ is unitary, we have
\begin{align}
1 = C_S C_S^\dagger = U^T \left( \begin{array}{cccc}
		c_{1}^{2} & & &\\ & c_{2}^{2} & & \\ & & \ddots & \\ & & & c_{N}^{2} \end{array} \right)
		U^{*} \,,
\end{align}
and hence $c_{1}^{2} = c_{2}^{2} = \cdots = c_{N}^{2} = 1$. Because all $c_i$ are positive-semi-definite, they all have to be one. Therefore, one can always choose a basis such that $C_S$ is just a phase factor, $C_{S}=e^{i\theta_S}$. In this basis, we clearly see from \cref{eqn:Ctransform} that the elements in $SU(N)$ that commute with $C_S$ satisfy
\begin{equation}
e^{i\theta_S} = C'_S = U^* C_S U^\dagger = U^* e^{i\theta_S} U^\dagger   \quad\Rightarrow\quad   U^T U = 1 \,,
\end{equation}
which form the $SO(N)$ subgroup in $SU(N)$. We can also choose an alternative basis where
\begin{align}
C_S = e^{i\theta_S} \left( \begin{array}{ccccc}
	& & & & 1 \\ & 0 & & 1 & \\ & & \iddots & & \\ & 1 & & 0 & \\ 1 & & & &
    \end{array} \right) \,,
\end{align}
which is more useful to identify the action of outer automorphism that reverses the Dynkin diagram as an action on roots and weights.

When $C=C_A$ is anti-symmetric, \emph{special} unitary transformations can always make it into skew-diagonal matrix with positive semi-definite eigenvalues $c_{i}$, up to a phase factor:
\begin{align}
C_{A} = U^{T} e^{i\theta_A} \left( \begin{array}{ccccc}
	0 & c_{1} & & & \\ -c_{1}& 0 & & & \\ & & \ddots & & \\
	& & & 0 & c_{k} \\ & & & -c_{k} & 0 \end{array} \right) U \,.
\end{align}
Of course, this is possible only when $N=2k$ is even. When $N$ is odd, the last eigenvalue has to be zero and $C_{A}$ cannot be unitary, consistent with the conclusion in the previous section. Now assuming $N=2k$, the unitarity of $C_{A}$ implies
\begin{align}
1 = C_{A} C_{A}^{\dagger}
= U^{T} \left( \begin{array}{ccccc}
	c_{1}^{2} & & & &\\ & c_{1}^{2} & & & \\ & & \ddots & & \\
	& & & c_{k}^{2} & \\ & & & & c_{k}^{2} \end{array} \right)
	U^{*} \,,
\end{align}
and hence $c_{1}^{2} = \cdots = c_{k}^{2} = 1$. Again because $c_{i}$ are positive-semi-definite, they all have to be one. Therefore, we can always choose a basis such that $C_A$ is the symplectic matrix $J$ multiplied by a phase factor:
\begin{align}
C_{A} = e^{i\theta_A} \left( \begin{array}{ccccc}
	0 & 1& & & \\ -1 & 0 & & & \\ & & \ddots & & \\
	& & & 0 & 1 \\ & & & -1& 0 \end{array} \right) = e^{i\theta_A} J \,.
\end{align}
In this basis, we see again from \cref{eqn:Ctransform} that elements in $SU(2k)$ that commute with $C_A$ satisfy
\begin{equation}
e^{i\theta_A} J = C'_A = U^* C_A U^\dagger = U^* e^{i\theta_A} J U^\dagger   \quad\Rightarrow\quad   U^T J U = J \,,
\end{equation}
which form the $Sp(2k)$ subgroup in $SU(2k)$. We can also choose an alternative basis where
\begin{align}
	C_{A} = e^{i\theta_A} \left( \begin{array}{ccc|ccc}
		& & & & & 1 \\ & 0 & & & \iddots & \\ & & & 1 & & \\ \hline
		& & -1 & & & \\ & \iddots & & & 0 & \\ -1 & & & & &
		\end{array} \right) \,,
\end{align}
which is more useful to identify the action of outer automorphism that reverses the Dynkin diagram as an action on roots and weights.

From the above discussion, we see that $\CC_S$ and $\CC_A$ are inequivalent charge conjugations under the criterion 3 in \cref{sec:general}, namely that we cannot find an element $g\in SU(2k)$ such that $\CC_A=g\,\CC_S\,g^{-1}$. As explained in \cref{sec:general}, in this case they are not the same physical observable, and they could also lead to different $\CC$-invariant subgroups. This is indeed verified in the above---the $\CC_S$-invariant subgroup of $SU(2k)$ is $SO(2k)$, while the $\CC_A$-invariant subgroup of $SU(2k)$ is $Sp(2k)$.

In fact, $\CC_S$ and $\CC_A$ are also inequivalent under criterion 2 in \cref{sec:general}, namely that we cannot find an element $g\in SU(2k)$ such that $\CC_A=g\,\CC_S$. To see this, let us assume such an $g$ exists. Then it must come with the form
\begin{equation}
g = \mqty(\Omega & 0 \\ 0 & \Omega^*) \,,
\label{eq:equivalence}
\end{equation}
with $\Omega$ an $N$ by $N$ special unitary matrix $\Omega^\dagger \Omega = 1$, $\det\Omega=1$. This would mean
\begin{equation}
\CC_A = g\,\CC_S   \quad\Rightarrow\quad
\left\{\begin{array}{rl}
C_A^\dagger &= \Omega\, C_S^\dagger \\[3pt]
C_A &= \Omega^*\, C_S
\end{array}\right. \,,
\end{equation}
which leads to the contradiction
\begin{equation}
1 = C_A^\dagger\, C_A = - C_A^\dagger\, C_A^T = - \Omega\, C_S^\dagger \left(\Omega^*\, C_S\right)^T = - \Omega\, C_S^\dagger\, C_S\, \Omega^\dagger = -1 \,.
\label{eqn:signproblem}
\end{equation}
Therefore, such a $g$ must not exist, which means that $\CC_S$ and $\CC_A$ are inequivalent under criterion 2 in \cref{sec:general}. A consequence of this is that $SU(2k)\rtimes \CC_S$ and $SU(2k)\rtimes \CC_A$ are actually different groups that are not isomorphic to each other. In this case, they could have different anomaly properties.

In fact, $\CC_S$ and $\CC_A$ are almost gauge equivalent, except that we also need a sign flip of either ${\bf N}$ or $\overline{\bf N}$, as implied by \cref{eqn:signproblem} above. More concretely, take the by-now-standard choice $C_S = I$, $C_A = J$. Using the gauge transformation
\begin{equation}
g = \mqty(J & 0 \\ 0 & J^*) \,,
\end{equation}
the transformation law is
\begin{align}
\left( \begin{array}{c} {\bf N} \\ \overline{\bf N} \end{array} \right)
\stackrel{{\cal C}_A}{\longrightarrow}
\left( \begin{array}{c} J^{-1} \overline{\bf N} \\ J {\bf N} \end{array} \right)
\stackrel{-g}{\longrightarrow}
\left( \begin{array}{c} -\overline{\bf N} \\ {\bf N} \end{array} \right)
\stackrel{{\cal C}_S}{\longrightarrow}
\left( \begin{array}{c} {\bf N} \\ -\overline{\bf N} \end{array} \right) \,,
\end{align}
which yields a state with a sign flip of $\overline{\bf N}$ but not ${\bf N}$. This additional sign flip can be identified as an element in the flavor transformation group $U(N_f)$. In other words, although $SU(2k)\rtimes \CC_S \ne SU(2k)\rtimes \CC_A$ as we have proved above in \cref{eqn:signproblem}, if one also includes the flavor transformation group $U(N_f)$, we actually have $\left[ SU(2k)\times U(N_f) \right] \rtimes \CC_S = \left[ SU(2k)\times U(N_f) \right] \rtimes \CC_A$. Therefore, the difference in the anomaly properties between $\CC_S$ and $\CC_A$ can be explained by the anomaly property of the flavor group $U(N_f)$. In particular, $\CC_S$ and $\CC_A$ should have the same anomaly properties when $N_f$ is even. We will discuss this point further in \cref{sec:para}.

\subsection{Self-conjugate Representations}

There are representations of $SU(N)$ self-conjugate under charge conjugation.  They are real or pseudo-real.  When $N$ is odd, they are obtained by tensor products of the same number of fundamental and anti-fundamental representations.  In general, they are given in tensor notation
\begin{align}
A^{i_{1} i_{2} \cdots i_{n}}_{j_{1} j_{2} \cdots j_{n}} \,,
\end{align}
where the symmetry properties among the indices are the same between upper and lower indices.  The charge conjugation acts on this representation as
\begin{align}
A^{i_{1} i_{2} \cdots i_{n}}_{j_{1} j_{2} \cdots j_{n}}
\rightarrow
\pm (C^{\dagger})^{i_{1}k_{1}}(C^{\dagger})^{i_{2}k_{2}}\cdots (C^{\dagger})^{i_{n}k_{n}}
C_{j_{1}l_{1}}C_{j_{2}l_{2}}\cdots C_{j_{n}l_{n}}
A^{l_{1} l_{2} \cdots l_{n}}_{k_{1} k_{2} \cdots k_{n}} \,.
\end{align}
The overall sign is the ``intrinsic charge conjugation'' for a given field.  Otherwise everything is fixed by the choice of the $C$ matrix.  Specifically for the adjoint representation, we can write the transformation as
\begin{align}
A^{i}{}_{j}
\rightarrow \pm (C^{\dagger})^{ik}
C_{jl}
A^{l}{}_{k} \,,
\end{align}
which can be written in the matrix notation
\begin{align}
A \rightarrow \pm C^{\dagger} A^{T} C^{T} \,.
\label{eqn:signassignment}
\end{align}
When the adjoint representation is identified as the generators, the sign above must be $-$ to preserve the Lie algebra. For a general field in the adjoint representation, however, there can be an intrinsic sign assignment $\pm$.

When $N=2k$ is even, there is another possibility of self-conjugate representations with rank-$k$ anti-symmetric tensor ${\cal A}_k$ (real for even $k$ and pseudo-real for odd $k$) and its tensor products with other real or pseudo-real representations. Using the Levi--Civita symbols,
\begin{align}
{\cal A}_k =
A^{i_{1} \cdots i_{k}}
\rightarrow \pm \frac{i^{k}}{k!} \epsilon^{i_{1} \cdots i_{k}j_{1}\cdots j_{k}}
C_{j_{1}l_{1}}\cdots C_{j_{k}l_{k}}
A^{l_{1} \cdots l_{k}} \,.
\end{align}
Here, the factor $\pm i^{k}$ makes sure that doing charge conjugation twice returns the original representation. In general, once the $C$ matrix is chosen for the fundamental and anti-fundamental representations, it determines the action of charge conjugation for any self-conjugate representations up to the intrinsic signs.

Now let us examine the relationship between ${\cal C}_S$ and ${\cal C}_A$ in self-conjugate representations when $N=2k$ is even. Similar to the case of fundamental and anti-fundamental representations, we will see that ${\cal C}_S$ and ${\cal C}_A$ are not guaranteed to be gauge equivalent, but they are equivalent (in the sense of our criterion 2) upon a further flavor transformation.

We start with the adjoint representation. Under ${\cal C}_S$, the generators transform as
\begin{align}
T_a \stackrel{{\cal C}_S}{\longrightarrow} - T_a^T \,,
\end{align}
while under ${\cal C}_A$,
\begin{align}
T_a \stackrel{{\cal C}_A}{\longrightarrow} - J T_a^T J^{-1} \,.
\end{align}
On the other hand, under the gauge transformation $J$,
\begin{align}
T_a \stackrel{J}{\longrightarrow} J^{-1} T_a J \,.
\end{align}
Performing all of the above, we find
\begin{align}
T_a \stackrel{{\cal C}_S}{\longrightarrow} - T_a^T
\stackrel{{\cal C}_A}{\longrightarrow} J T_a J^{-1}
\stackrel{J}{\longrightarrow} T_a \,.
\end{align}
Therefore, ${\cal C}_S$ and ${\cal C}_A$ are gauge equivalent in the adjoint representation up to the intrinsic sign assignment in \cref{eqn:signassignment}. However, this sign assignment can be identified as a flavor transformation.

Going beyond the adjoint representation, we make use of the fact that any self-conjugate representation can be obtained by an appropriate tensor product of ${\cal A}_k$. Taking the gauge element
\begin{align}
g = \text{diag} ( \underbrace{i, \cdots, i}_{k}, \underbrace{-i, \cdots, -i}_{k}) \in SU(2k) \,,
\end{align}
one can check that in the ${\cal A}_k$ representation, ${\cal C}_S$ is gauge equivalent to ${\cal C}_A$ up to an overall factor of $\pm i^k$, which can again be identified with a flavor transformation. This completes the proof that in any self-conjugate representation ${\cal C}_S$ and ${\cal C}_A$ are related by a gauge and a flavor transformation.

\subsection{$\CC$-invariant Subgroups}

When we discuss anomalies that involve charge conjugation, we need to figure out how the fermion path integral measures transform under the charge conjugation.  There is one important problem: the charge conjugation does not commute with the gauge symmetry precisely due to its definition being an outer automorphism.  The charge conjugation may not commute with other global symmetries either. So it is necessary to study the subgroups of gauge and global symmetries that commute with the charge conjugation.  Here we summarize some of the important cases relevant for later discussions.

We have seen that the charge conjugation $\CC_{S}$ of an $SU(N)$ group leaves an $SO(N)$ subgroup invariant.  For the fundamental ${\bf N}$ and anti-fundamental $\overline{\bf N}$ representations, they become identical vector representations under the $\CC_S$-invariant subgroup, and they are interchanged under the charge conjugation.  Therefore, the two linear combinations ${\bf N}\pm \overline{\bf N}$ are ``eigenstates'' under the charge conjugation with definite signs as ``eigenvalues.''  This property is extended to higher representations as well.  For instance, anti-symmetric tensors $A$ of $SU(N)$ are irreducible anti-symmetric tensors of $SO(N)$.  $A$ and $A^{*}$ are interchanged and we again find two eigenstates with opposite signs.  There is one exception to this rule when $N=2k$ is even.  The rank $k$ anti-symmetric tensor decomposes as ``self-dual'' and ``anti-self-dual'' rank $k$ tensors\footnote{As we discussed earlier, the Hodge dual performed twice is trivial $*(*\omega)=\omega$ when $k$ is even, and the self-dual is defined by $*\omega_{k} = \omega_{k}$ while the anti-self-dual is $*\omega_{k}=-\omega_{k}$.  On the other hand, when $k$ is odd, $*(*\omega)=-\omega$.  Therefore self-dual and anti-self-dual means $*\omega = \pm i \omega$.  This point is also related to the fact that the rank-$k$ anti-symmetric tensor is real for $k$ even and pseudo-real for $k$ odd.} of $SO(2k)$.  We assign opposite signs for them under charge conjugation.
For self-conjugate representations, they decompose into several representations under the $\CC$-invariant subgroups.  For example, an adjoint representation of $SU(N)$ decomposes into a traceless symmetric tensor $\Ysymm$ and an anti-symmetric tensor $\Yasymm$ representation of $SO(N)$, and they transform with opposite signs under the charge conjugation.

The same consideration applies to global symmetries.  For example, QCD-like theories based on $SU(N)$ gauge groups have $SU(F)_{L} \times SU(F)_{R}$ global flavor symmetry.  Under charge conjugation, $L$ and $R$ switches, and the $\CC$-invariant subgroup is the diagonal subgroup $SU(F)_{C}$.  Note that this is {\it not}\/ the usual $SU(F)_{V}$ where $V_{L} = V_{R}$.  To correctly identify the $\CC$-invariant subgroup of the flavor symmetry, we need to treat the relevant fields on an equal footing.  Therefore we need to take the charge conjugation\footnote{This is the charge conjugation in textbooks for Dirac fields $\psi \rightarrow \psi^{c} = -i \gamma^{0}\gamma^{2} \bar{\psi}^{T}$.} of the right-handed quarks and treat them as left-handed anti-quarks in the anti-fundamental representation of the color group.  Then under the $\CC$-invariant subgroup $SO(N)$ of the gauge group, quarks in ${\bf N}$ and anti-quarks in $\overline{\bf N}$ are interchanged under $\CC_{S}$.  The anti-quark field transforms by $V_{R}^{*}$, and the $\CC$-invariant subgroup $SU(F)_C$ is defined by $V_{L} = V_{R}^{*}$. The mesons decompose as a symmetric tensor $\Ysymm$ and an anti-symmetric tensor $\Yasymm$ representation of $SU(F)_C$, and they transform with opposite signs under the charge conjugation.

For the other charge conjugation $\CC_A$ for $SU(2k)$, we apply the same consideration to the $\CC$-invariant subgroup $Sp(2k)$.  We decompose each $SU(2k)$ representations to irreducible representations of $Sp(2k)$, and assign signs to each of them appropriately. For example, an adjoint representation of $SU(2k)$ decomposes into a symmetric tensor $\Ysymm$ and a traceless anti-symmetric tensor $\Yasymm$ representation of $Sp(2k)$, and they transform with opposite signs under the charge conjugation. As another example, the rank-$k$ antisymmetric tensor of $SU(2k)$ decomposes into $Sp(2k)$ as

{\tiny $$
\begin{ytableau}
 \none[\bullet]
\end{ytableau}~
\oplus~
\begin{ytableau}
 \\
 \\
\end{ytableau}
~\oplus~\begin{ytableau}
 \\
 \\
 \\
 \\
\end{ytableau}
~\oplus\cdots\oplus~
\begin{ytableau}
 \\
 \\
 \\
 \\
\none[\vdots] \\
\\
\\
\end{ytableau}
$$
}\noindent
for $k$ even (with the final column consisting of $k$ boxes) and
{\tiny $$
\begin{ytableau}
 \\
\end{ytableau}~
\oplus~
\begin{ytableau}
 \\
 \\
 \\
\end{ytableau}
~\oplus~\begin{ytableau}
 \\
 \\
 \\
 \\
 \\
\end{ytableau}
~\oplus\cdots\oplus~
\begin{ytableau}
 \\
 \\
 \\
 \\
  \\
\none[\vdots] \\
\\
\\
\end{ytableau}
$$
}\noindent
for $k$ odd (with the final column again consisting of $k$ boxes); the representations in the above sums transform with alternating sign under charge conjugation.

\section{Parities for General \boldmath$SO(2r)$}
\label{sec:parity}

As discussed in \cref{sec:general}, $SO(2r)$ in general can have ``parity'' defined in a way to break the symmetry to $SO(q)\times SO(2r-q)$ with $q$ odd. Namely the parity is defined by an element of $O(2r)$,
\begin{align}
\PP_q = {\rm diag}(\underbrace{+, \cdots, +}_{q},\underbrace{-,\cdots,-}_{2r-q}) \,.
\end{align}

Take the example of $SO(6) \simeq SU(4)$. For ${\cal P}_{1}=(+, -, -, -, -, -)$, the $\PP$-invariant subgroup is $SO(5) \simeq Sp(4)$. This is consistent with the charge conjugation of $SU(4)$ with $\CC_{A}$. On the other hand, for ${\cal P}_{3}=(+, +, +, -, -, -)$, the $\PP$-invariant subgroup is $SO(3)\times SO(3) \simeq SO(4)$. This is consistent with the charge conjugation of $SU(4)$ with $\CC_{S}$.

For general $SO(2r)$, there are a wider variety of parities. We can study them by constructing an explicit representation of Spin($2r$)(the double cover of $SO(2r)$)---we let each $\gamma_i$ be a tensor product of $r$ Pauli matrices:
\begin{equation}\renewcommand\arraystretch{1.3}
\begin{array}{rccccccccc}
\gamma_1^{}      =& \sigma_1 & \otimes &        1 & \otimes & \cdots & \otimes &        1 & \otimes & 1 \,, \\
\gamma_2^{}      =& \sigma_2 & \otimes &        1 & \otimes & \cdots & \otimes &        1 & \otimes & 1 \,, \\
\gamma_3^{}      =& \sigma_3 & \otimes & \sigma_1 & \otimes & \cdots & \otimes &        1 & \otimes & 1 \,, \\
\gamma_4^{}      =& \sigma_3 & \otimes & \sigma_2 & \otimes & \cdots & \otimes &        1 & \otimes & 1 \,, \\
        \vdots &&&&&&&&& \\
\gamma_{2r-3}^{} =& \sigma_3 & \otimes & \sigma_3 & \otimes & \cdots & \otimes & \sigma_1 & \otimes & 1 \,, \\
\gamma_{2r-2}^{} =& \sigma_3 & \otimes & \sigma_3 & \otimes & \cdots & \otimes & \sigma_2 & \otimes & 1 \,, \\
\gamma_{2r-1}^{} =& \sigma_3 & \otimes & \sigma_3 & \otimes & \cdots & \otimes & \sigma_3 & \otimes & \sigma_1 \,, \\
  \gamma_{2r}^{} =& \sigma_3 & \otimes & \sigma_3 & \otimes & \cdots & \otimes & \sigma_3 & \otimes & \sigma_2 \,, \\
\gamma_{2r+1}^{} =& \sigma_3 & \otimes & \sigma_3 & \otimes & \cdots & \otimes & \sigma_3 & \otimes & \sigma_3 \,. \\
\end{array}
\label{eq:gammas}
\end{equation}
The last one $\gamma_{2r+1}^{}$ plays the role of $\gamma_5$ in case of $2r=4$ --- one can readily verify
\begin{equation}
\gamma_{2r+1}^{} = (-i)^r \gamma_1^{}\, \gamma_2^{}\, \cdots \gamma_{2r-1}^{}\, \gamma_{2r}^{} \,.
\label{eqn:gamma5}
\end{equation}
It is easy to check the gamma matrices in \cref{eq:gammas} satisfy the Clifford algebra $\{\gamma_{i}^{}, \gamma_{j}^{}\} = 2 \delta_{ij}$ for Spin($2r$). The $\mathfrak{so}(2r)$ Lie algebra is then generated by
\begin{align}
\frac{1}{2} \sigma_{ij} = \frac{i}{4} [ \gamma_{i}^{}, \gamma_{j}^{}] \,.
\end{align}
Spin($2r$) groups extended to include the parity are called Pin($2r$), removing ``S'' that stands for ``Special'' for unit determinant to allow for determinant of $-1$, as a joke attributed to Jean-Pierre Serre.

In this convention, all $\gamma_{2n-1}^{}$ are symmetric while $\gamma_{2n}^{}$ are anti-symmetric. It is natural to separate them into two separate groups and define parities accordingly. In what follows, we show this explicitly for the cases of even $r=2k$ and odd $r=2k+1$ separately.

\subsection{Even $r=2k$}

In this case, all spinor representations are real ($k=2\ell$) or pseudo-real ($k=2\ell+1$). The parities can be defined by the action of
\begin{align}\renewcommand\arraystretch{1.5}
	\begin{array}{cclcccl}
	{\cal P}_{1} &=& \gamma_{1}^{} \,, & \qquad\qquad
		& \bar{\cal P}_{1} &=& i {\cal P}_{1} \gamma_{2r+1}^{} \,, \\
	{\cal P}_{3} &=& i \gamma_{1}^{} \gamma_{3}^{} \gamma_{5}^{} \,, & \qquad\qquad
		& \bar{\cal P}_{3} &=& i {\cal P}_{3} \gamma_{2r+1}^{} \,, \\
	& \vdots & &\qquad\qquad & &\vdots & \\
	{\cal P}_{2n-1} &=& i^{n-1} \gamma_{1}^{} \gamma_{3}^{} \cdots \gamma_{4n-3}^{} \,, & \qquad\qquad
		& \bar{\cal P}_{2n-1} &=& i {\cal P}_{2n-1} \gamma_{2r+1}^{} \,, \\
	& \vdots & &\qquad\qquad & &\vdots & \\
	{\cal P}_{2k-1} &=& i^{k-1} \gamma_{1}^{} \gamma_{3}^{} \cdots \gamma_{4k-3}^{} \,,
		& \qquad\qquad & \bar{\cal P}_{2k-1} &=& i {\cal P}_{2k-1} \gamma_{2r+1}^{} \,.
	\end{array}
\end{align}
The definitions above make sure that they all satisfy $\PP^2 = 1$ and $\PP^\dagger \PP = 1$.

Note that although $\PP_{2n-1}$ and $\bar{\PP}_{2n-1}$ lead to the same $\PP$-invariant subgroup $SO(2n-1) \times SO(2r-2n+1)$, they have
different symmetry properties:
\begin{equation}
\PP_{2n-1}^T = (-1)^{n-1}\, \PP_{2n-1} \,,\qquad   \bar{\PP}_{2n-1}^T = (-1)^n\, \bar{\PP}_{2n-1} \,.
\end{equation}
In particular, $\PP_{2n-1}$ is a product of $(2n-1)$ gamma matrices, while $\bar{\PP}_{2n-1}$ is a product of $2r-(2n-1)$ gamma matrices. Since the number of gamma matrices is preserved by Spin($2r$) basis changes, all of them above are inequivalent under the criterion 3 in \cref{sec:general}.

How about criterion 2 in \cref{sec:general}, namely $\PP_A \simeq \PP_B \;\Leftrightarrow\; \PP_B = g \PP_A$ for $g\in\text{Spin}(2r)$?
To see the equivalence relations among these parities under criterion 2, we recall that the Spin($2r$) generators are
\begin{equation}
\frac12 \sigma_{ij} = \frac12 i\gamma_i^{} \gamma_j^{} \,,
\end{equation}
and in particular we have the elements
\begin{equation}
\exp\left(\pm i\pi \frac12 \sigma_{ij} \right) = \mp \gamma_i^{} \gamma_j^{} \,.
\end{equation}
Any products of these elements are also Spin($2r$) elements. With these, we can easily see that the following four sets each forms an equivalence class under criterion 2:
\begin{subequations}
\begin{alignat}{2}
\PP_1 \,&\simeq\, \PP_5 \,\simeq\, \PP_9 \,\simeq\, \cdots \;,\qquad&\qquad
\bar{\PP}_1 \,&\simeq\, \bar{\PP}_5 \,\simeq\, \bar{\PP}_9 \,\simeq\, \cdots \,, \\[5pt]
\PP_3 \,&\simeq\, \PP_7 \,\simeq\, \PP_{11} \,\simeq\, \cdots \;,\qquad&\qquad
\bar{\PP}_3 \,&\simeq\, \bar{\PP}_7 \,\simeq\, \bar{\PP}_{11} \,\simeq\, \cdots \,.
\end{alignat}
\end{subequations}
However, they are still not all inequivalent. In the case $r=2k$ even, there are an even number of $-i$ factors in $\gamma_{2r+1}$ (see \cref{eqn:gamma5}) and one can further verify that
\begin{subequations}\label{eqn:PPequivalenceEvenr}
\begin{align}
\PP_1 \,&\simeq\, \bar{\PP}_3 \,, \\[5pt]
\PP_3 \,&\simeq\, \bar{\PP}_1 \,.
\end{align}
\end{subequations}
So they reduce to two equivalence classes, represented by $\PP_1$ and $\PP_3$ respectively. To see that these two are indeed inequivalent, we construct the following matrix $\Gamma$, which transforms the Spin($2r$) generators as
\begin{equation}
\Gamma \equiv \gamma_2^{}\, \gamma_4^{}\, \cdots\, \gamma_{2r-2}^{}\, \gamma_{2r}^{}
\qquad\Rightarrow\qquad
\Gamma\, \sigma_{ij}\, \Gamma^{-1} = -\sigma_{ij}^T \,.
\end{equation}
This tells us that any element $g \in \text{Spin}(2r)$ must satisfy
\begin{equation}
g^T\, \Gamma\, g = \Gamma \,, \qquad \forall\, g \in \text{Spin}(2r) \,.
\label{eqn:spinelement}
\end{equation}
On the other hand
\begin{equation}
\PP_1 \PP_3 = i \gamma_3^{} \gamma_5^{} \,,
\label{eqn:P1P3}
\end{equation}
does not satisfy \cref{eqn:spinelement}. So $\PP_1 \PP_3$ is not a Spin($2r$) element, namely that the two parities cannot be linked by a multiplication of a Spin($2r$) element. Therefore, we end up with two inequivalent classes of parities under criterion 2 in \cref{sec:general};
they are the analogs of $\CC_S$ and $\CC_A$ that we saw in the case of $SU(2k)$.

\subsection{Odd $r=2k+1$}

In this case, all spinor representations are complex. The parities can be defined by the action of
\begin{align}\renewcommand\arraystretch{1.5}
	\begin{array}{cclcccl}
	{\cal P}_{1} &=& \gamma_{1}^{} \,, & \qquad\qquad
		& \bar{\cal P}_{1} &=& i {\cal P}_{1} \gamma_{2r+1}^{} \,, \\
	{\cal P}_{3} &=& i \gamma_{1}^{} \gamma_{3}^{} \gamma_{5}^{} \,, & \qquad\qquad
		& \bar{\cal P}_{3} &=& i {\cal P}_{3} \gamma_{2r+1}^{} \,, \\
	& \vdots & &\qquad\qquad & &\vdots & \\
	{\cal P}_{2n-1} &=& i^{n-1} \gamma_{1}^{} \gamma_{3}^{} \cdots \gamma_{4n-3}^{} \,, & \qquad\qquad
		& \bar{\cal P}_{2n-1} &=& i {\cal P}_{2n-1} \gamma_{2r+1}^{} \,, \\
	& \vdots & &\qquad\qquad & &\vdots & \\
	{\cal P}_{2k+1} &=& i^k \gamma_{1}^{} \gamma_{3}^{} \cdots \gamma_{4k+1}^{} \,,
		& \qquad\qquad & \bar{\cal P}_{2k+1} &=& i {\cal P}_{2k+1} \gamma_{2r+1}^{} \,.
	\end{array}
\end{align}
What is different from the case $r=2k$ is that now $\gamma_{2r+1}$ has an odd number of $-i$ factors in it (see \cref{eqn:gamma5}). Consequently, \cref{eqn:PPequivalenceEvenr} now needs to be replaced by
\begin{subequations}\label{eqn:PPequivalenceOddr}
\begin{align}
\PP_1 \,&\simeq\, \bar{\PP}_1 \,, \\[5pt]
\PP_3 \,&\simeq\, \bar{\PP}_3 \,.
\end{align}
\end{subequations}
Since the argument around \cref{eqn:P1P3} still holds, there are still two inequivalent classes of parities under criterion 2 in \cref{sec:general}.

\subsection{Relation to the pin  group}

The two equivalence classes above with representatives ${\cal P}_1$ and  ${\cal P}_3$ correspond to the groups Pin$_\pm$. In the above we only consider the case where they square to $+1$. If we further allow for the parity to square to $-1$, then there are more parities, corresponding to a factor of $i$ multiplying all of the above ${\cal P}_i$. In this case, we find the following equivalences under multiplication of a Spin$(n)$ element: ${\cal P}_1\simeq i {\cal P}_3\simeq {\cal P}_5\simeq \ldots$, and $i{\cal P}_1\simeq  {\cal P}_3\simeq i{\cal P}_5 \simeq \ldots$. Thus the equivalence classes can be represented by ${\cal P}_1$ and $i{\cal P}_1$, whence the identification with Pin$_\pm$. Therefore, the two parities are gauge equivalent up to an overall factor of $\pm i$ on the field and have different anomalies in general.

\subsection{Decompositions under $\PP$-invariant Subgroups}

The parity for $SO(2r)$ were already studied in the literature concerning discrete anomalies \cite{Csaki:1997aw}. However, it was studied only in the context that the mapping between discrete symmetries between the electric and magnetic dual theories sometimes include the parity. The anomaly associated with parity itself was not studied.

In the rest of this paper, we will focus on $\PP_1$ when discussing anomalies involving parity. The $\PP_1$-invariant subgroup of $SO(2r)$ is $SO(2r-1)$. Under this subgroup, a vector representation decomposes as ${\bf 2r}={\bf 2r-1} \oplus {\bf 1}$, where the singlet ${\bf 1}$ switches its sign. An adjoint representation decomposes as ${\bf adj}_{2r} = {\bf adj}_{2r-1} \oplus {\bf 2r-1}$, where the vector ${\bf 2r-1}$ switches its sign. On top of this, one can assign an ``intrinsic parity'' on each field.

\section{\boldmath$E_6$ and $SO(8)$}
\label{sec:E6}

As discussed in \cref{sec:general}, there are two outer automorphisms for $E_{6}$. One of them has the $\CC$-invariant subgroup $Sp(8)$, where the $E_6$ fundamental representation ${\bf 27}$ becomes $Sp(8)$ rank-two anti-symmetric representation
\begin{equation}
{\bf 27} = \frac12 (8\times 7) - 1 \,.
\end{equation}
The $E_6$ adjoint representation ${\bf 78}$ decomposes as
\begin{subequations}
\begin{align}
{\bf 36} &= \Ysymm = \frac12 (8 \times 9) \in ({\bf 27} \otimes {\bf 27})_{A} \,,\\[5pt]
{\bf 42} &= \Yfoura \in ({\bf 27} \otimes {\bf 27})_{S} \,.
\end{align}
\end{subequations}
The ${\bf 36}$ is even under the charge conjugation and anti-symmetric under the interchange of two ${\bf 27}$, while the remaining ${\bf 42}$ is odd under the charge conjugation and symmetric under the interchange of two ${\bf 27}$. This is the analog of $\CC_S$ that we saw in the case of $SU(2k)$.

The other outer automorphism of $E_6$ has the invariant subgroup $F_{4}$, where the fundamental representation ${\bf 27}$ is decomposes as ${\bf 26} \oplus {\bf 1}$. The adjoint representation ${\bf 78}$ decomposes as ${\bf 52} \oplus {\bf 26}$. Here, ${\bf 52} \in ({\bf 26} \otimes {\bf 26})_{A}$ is the adjoint of $F_{4}$ which is even under charge conjugation, while ${\bf 26}$ is odd. Namely the $E_{6}$ generators can be viewed as $27\times 27$ matrix where the charge conjugation is given by the diagonal matrix
\begin{align}
	(+,\underbrace{-,-,\cdots,-,-}_{26}),
\end{align}
similar to that of the parity $\PP_1$ for $SO(2r)$. This is an analog of the outer automorphism $\CC_A$ that we saw in the case of $SU(2k)$.

The two outer automorphisms are inequivalent under the criterion 2. in Sec.~\ref{sec:general}. However, just like in the case of $SU(2k)$, they are related by a combination of gauge and flavor transformations. To identify the combination, we look at the subgroup $Sp(2)\times Sp(6)$ common between $Sp(8)$ and $F_4$. The adjoint representation decomposes as
\begin{align}
	{\bf 78} = {\bf 28} (\Yfund, \Ythreea) \oplus {\bf 14} (\cdot, \Yasymm) \oplus
		{\bf 21} (\cdot, \Ysymm) \oplus {\bf 12} (\Yfund, \Yfund) \oplus {\bf 3} (\Ysymm, \cdot).
\end{align}
In addition to ${\cal C}_S$ and ${\cal C}_A$, we consider ${\mathbb Z}_2$ center of $Sp(2)$ as the gauge transformation. The multiplets transform as
\begin{align}
	\begin{tabular}{|c||c|c|c|c|c|} \hline
	& {\bf 28} & {\bf 14} & {\bf 21} & {\bf 12} & {\bf 3} \\ \hline
	$Sp(8)$ & $-$ & $-$ & $+$ & $+$ & $+$ \\
	$F_4$ & $+$ & $-$ & $+$ & $-$ & $+$ \\
	gauge & $-$ & $+$ & $+$ & $-$ & $+$ \\ \hline
	\end{tabular}
\end{align}
The product of all three is $+$ for all multiplets, and hence ${\cal C}_S$ and ${\cal C}_A$ are gauge equivalent, while an intrinsic charge conjugation is possible with $\pm 1$.

In the case of $SO(8)$, there is a very special {\it triality}\/ $S_3$ outer automorphism as discussed in \cref{sec:general}. It interchanges the vector and two inequivalent spinor representations. The $\CC$-invariant subgroup is $G_{2}$ under which the vector and spinor representations decompose as ${\bf 8}={\bf 7}\oplus {\bf 1}$ while the adjoint as ${\bf 28} = {\bf 14} \oplus {\bf 7} \oplus {\bf 7}$.

For the outer automorphisms of $E_6$ and the triality of $SO(8)$, we are not aware of well-established examples of gauge theories where we can study their anomalies.

\section{A Paradox about Charge Conjugation in QCD?}
\label{sec:para}

The massless QCD Lagrangian is invariant under charge conjugation.  To make this point most apparent, we write $N_{f}$ quark fields as left-handed Weyl fermions, where $q_{i}$ is in the fundamental ${\bf N}$ representation, while $\tilde{q}_{i} = (q_{i R})^{c}$ is in the anti-fundamental $\overline{\bf N}$ representation, where $i=1, \cdots, N_{f}$.  Then the Lagrangian is
\begin{align}
{\cal L} &= -\frac{1}{2} {\rm Tr} G_{\mu\nu} G^{\mu\nu}
+ \left(\begin{array}{cc}\bar{q}_{i} & \bar{\tilde{q}}_{i} \end{array} \right)
\left( \begin{array}{cc} i \gamma^{\mu} (\partial_{\mu} - i g A_{\mu}) & 0 \\
	0 & i \gamma^{\mu} (\partial_{\mu} + i g A_{\mu}^{T}) \end{array} \right)
\left( \begin{array}{c} q_{i} \\ \tilde{q}_{i} \end{array} \right) \,.
\end{align}
Under the charge conjugation \cref{eq:CgC}, or more specifically
\begin{subequations}
\begin{align}
q &\to C^\dagger \tilde{q} \,, \\
\tilde {q} &\to C q \,,
\end{align}
\end{subequations}
it is guaranteed that this Lagrangian is invariant, if the gauge field is also transformed as
\begin{align}
A_{\mu} \rightarrow - C^\dagger A_{\mu}^{T} C \,.
\label{eqn:Atransform}
\end{align}
Note that this is the {\it classical}\/ invariance of the Lagrangian.  It remains to be seen whether it is respected at the quantum level, or in other words, whether the charge conjugation is anomalous.

We all believe that QCD-like gauge theories with relatively few number of flavors lead to dynamical chiral symmetry breaking with a chiral condensate
\begin{align}
\langle (\tilde{q}^{T} q) \rangle \propto {\bf 1}_{N_{f}} \neq 0 \,.
\end{align}
Under the charge conjugation, the quark bilinear transforms as
\begin{align}
\tilde{q}^{T} q \rightarrow (C q)^{T} (C^{\dagger} \tilde{q})
= (\tilde{q}^{T} C^{T\dagger}) (C q)
= \pm \tilde{q}^{T} q \,,
\end{align}
where the sign depends on $C^{T} = \pm C$.  Here, we implicitly used the fact that the spinor indices are contracted with the anti-symmetric tensor $\epsilon_{\alpha\beta}$ while the fermion fields anti-commute, so that we can treat $q$ and $\tilde{q}$ as if they commute.

It is a paradox that the chiral condensate is forbidden under the anti-symmetric charge conjugation $\CC_A$. For instance, consider an $SU(4)$ gauge theory with three flavors of quarks. In this theory, baryons are bosons so we do not have any candidate massless fermions to match the $U(1)_{B} (SU(3)_L)^2$ anomaly, which means this symmetry must be spontaneously broken in the IR. Note that the $U(1)_B$ symmetry is not spontaneously broken due to the Vafa--Witten theorem~\cite{Vafa:1983tf}, so the chiral symmetry must be broken. Yet the quark bilinear condensate appears forbidden by the charge conjugation. It appears to require that the charge conjugation symmetry is also broken spontaneously by the chiral condensate, which leads to two discrete sets of vacua (each with the coset space $G/H$) and domain walls. This is very unlikely. For instance, it does not happen in the class of SQCD with anomaly-mediated supersymmetry breaking analyzed in Refs.~\cite{Murayama:2021xfj, Csaki:2021xhi, Csaki:2021aqv}. We believe this paradox is resolved because the charge conjugation $\CC_A$ is anomalous.

The unambiguous way to study the anomaly is to look at the transformation of the Euclidean path integral measure under the symmetry.  For the usual chiral anomaly, the path integral measure for fermion can be decomposed under the eigenmodes of the Dirac operator
\begin{subequations}
\begin{align}
i\slashed{D}\, \psi_{nR} &= \lambda_{n}\, \psi_{nL} \,, \\[3pt]
i\slashed{D}\, \psi_{nL} &= \lambda_{n}\, \psi_{nR} \,,
\end{align}
\end{subequations}
which are always paired between left- and right-handed fermions when $\lambda_{n} \neq 0$.  But for zero modes $\lambda_{n}$ they do not need to be paired, and the mismatch is given by the index theorem
\begin{align}
n_{L}^{0} - n_{R}^{0} = \text{index} \left(i\slashed{D}\right)
= \frac{g^{2}}{32\pi^{2}} \int \epsilon_{\mu\nu\rho\sigma} \Tr\left(G^{\mu\nu} G^{\rho\sigma}\right)
= \#\mbox{ instantons} \,.
\end{align}
The path integral measure is then
\begin{align}
{\cal D} \psi_{L} {\cal D}\psi_{R}
= \prod_{n \neq 0} d \psi_{nL} d \psi_{nR}
\prod_{i=1}^{n_{L}^{0}} d \psi_{0L}^{i} \prod_{i=1}^{n_{R}^{0}} d \psi_{0R}^{i} \,.
\label{eq:Fujikawa}
\end{align}
Under the chiral transformation $\psi_{L} \rightarrow e^{-i\theta} \psi_{L}$, $\psi_{R} \rightarrow e^{+i\theta} \psi_{R}$, the path integral measure changes by $e^{-i (n_{L}^{0} - n_{R}^{0}) \theta}$ and hence is not invariant.  This is the origin of the chiral anomaly.

We can apply the same argument to discrete symmetries \cite{Csaki:1997aw, Araki:2006sqx}.  By considering an instanton background, a ${\mathbb Z}_{2}$ symmetry may change the sign of the path integral measure.  Then the ${\mathbb Z}_{2}$ symmetry is anomalous and hence is not a symmetry of the gauge theory. To apply this argument to charge conjugation, we must decompose the fermion fields into even and odd eigenstates under the charge conjugation.  However, this requires the instanton to have well-defined number of zero modes for even and odd states, and hence the instanton must belong to the $\CC$-invariant subgroup.

For the symmetric charge conjugation, the $\CC_S$-invariant subgroup is $SO(N)$.  The $q$ and $\tilde{q}$ fields are both in the vector representation of $SO(N)$ and they are not distinguished under the $SO(N)$ instanton background.  Because $q$ and $\tilde{q}$ are interchanged under the charge conjugation, the linear combinations $q \pm \tilde{q}$ are even and odd eigenstates.  Therefore there are $N_{f}$ even and $N_{f}$ odd eigenstates under the charge conjugation. The $N_f$ odd eigenstates $q - \tilde{q}$ could potentially make $\CC_S$ anomalous. However, since each flavor of $q - \tilde{q}$ forms the vector representation of $SO(N)$, which always has an even number of zero modes under the minimum $SO(N)$ instanton, the path integral measure is always invariant under $\CC_S$. Therefore, $\CC_S$ is always non-anomalous.\footnote{For the purpose of showing that a symmetry is anomalous, it is sufficient to pick a background gauge field under which the path integral measure is not invariant. For the purpose of showing that a symmetry is non-anomalous, however, we need to show that the path integral measure is invariant under {\it any}\/ background gauge fields. Since the argument here picks only background gauge fields that are invariant under the charge conjugation, it leaves some concern whether the invariance is indeed true. Yet the experimental fact that the charge conjugation is a symmetry of the strong interactions give us confidence. See \cref{ap:ap1} for further discussion.}
More concretely, for $N\ge4$, the vector representation of $SO(N)$ has {\it two}\/ zero modes under the minimum instanton, as $\tr_\Box\left(t^at^b\right)=2\cdot\frac12\delta^{ab}$. The easiest way to understand it is by breaking $SO(N)$ to $SU([\frac{N}{2}])$ where the vector representation decomposes as $[\frac{N}{2}] \oplus [\frac{N}{2}]^{*} (\oplus 1)$ (the singlet is there when $N$ is odd).  Since both $[\frac{N}{2}]$ and $[\frac{N}{2}]^{*}$ have one zero mode each, there are two zero modes altogether. For $N=3$, the vector representation has {\it four}\/ zero modes under the $SO(3)$ instanton, as $\tr\left(t^at^b\right)=4\cdot\frac12\delta^{ab}$ for the vector representation of $SO(3)$. For $N=2$, the $\CC$-invariant subgroup of $SU(2)$ under $\CC_S$ is $SO(2)\simeq U(1)$, and no instanton could be formed. Nevertheless, one may still be concerned with the $\CC_S U(1)^2$ anomaly. In this case, each flavor of $q - \tilde{q}$ forms the doublet representation of $SU(2)$, the two components of which share the same charge $q$ under the $\CC$-invariant subgroup $U(1)$. So its contribution to the $\CC_S U(1)^2$ anomaly is $2q^2$, always an even number.

For the anti-symmetric charge conjugation, the $\CC_A$-invariant subgroup is $Sp(N)$, possible only when $N$ is even. The $q$ and $\tilde{q}$ fields are both in the fundamental representation of $Sp(N)$ and they are not distinguished under the $Sp(N)$ instanton background.  Because $q$ and $\tilde{q}$ are interchanged under the charge conjugation, the linear combinations $q \pm \tilde{q}$ are even and odd eigenstates.  Therefore there are $N_f$ even and $N_f$ odd eigenstates under the charge conjugation. Unlike the case of $\CC_S$, the fundamental representation of $Sp(N)$ group has {\it one}\/ zero mode under its minimum instanton, as $\tr_\Box\left(t^at^b\right)=\frac12\delta^{ab}$ for $Sp(N)$. Therefore the path integral measure changes its sign as $(-1)^{N_f}$ under $\CC_A$. We conclude that $\CC_A$ is anomalous when $N_f$ is odd. For odd $N_f$, therefore, there is no charge conjugation invariance $\CC_A$, and hence the existence of chiral condensate is consistent without two discrete sets of vacua.

For even $N_f$ we have two charge conjugations $\CC_S$ and $\CC_A$ as symmetries of the theory. Under $\CC_S$, the chiral condensate is transposed, and it respects the charge conjugation because it is proportional to an identity matrix. On the other hand, under $\CC_A$, it still changes its sign. However, given that $N_f$ is even, the $SU(N_f)_L$ flavor symmetry has the $\mathbb{Z}_{N_f}$ center, which contains the element $-1$. So it is the $\CC_A$ compensated by this element that should be identified as the unbroken anti-symmetric charge conjugation symmetry.

Note that the observation in this section is consistent with the fact that $\CC_A$ is considered gauge equivalent to $\CC_S$ together with a sign flip of all $q$ (or $\bar{q}$ but not both). Each sign flip is anomalous, while the overall anomaly is indeed given by $(-1)^{N_f}$.

\section{Anomaly Matching}
\label{sec:anom}

Here we demonstrate the anomaly matching condition for charge conjugation as well as other outer automorphisms using well-established dualities in ${\cal N}=1$ SQCD.

\subsection{Review of Discrete Anomaly Matching}

In this section, we summarize main results of the paper by Csaba Cs\'aki and one of the authors (HM) \cite{Csaki:1997aw} that studied the anomaly matching conditions of discrete symmetries. Here we pay a special attention to ${\mathbb Z_{2}}$ symmetries. There are Type-I and Type-II anomalies.

The Type-I anomaly matching condition is uncontroversial. The anomaly arises when the fermion determinant is not invariant under the discrete symmetry. First we promote a continuous non-anomalous non-abelian global symmetry to a gauge symmetry. Then under an instanton background of this gauge symmetry, the Fujikawa measure of the fermion path integral is decomposed similar to \cref{eq:Fujikawa}. Now under a discrete symmetry $\mathbb{Z}_{N}$, the fermion field transforms as $\psi \rightarrow \omega \psi$ with $\omega^{N}=1$. If the fermion path integral is not invariant under this transformation, then there is an anomaly for the discrete symmetry. For an anomaly matching argument, two theories, either UV vs IR or electric vs magnetic, must behave the same way. This is the same as studying triangle anomalies by assigning a ``$U(1)$ charge'' under the $\mathbb{Z}_{N}$ symmetry, and the anomalies must match modulo $N$.

The argument can be applied also for the gravitational instanton background. It needs to be remembered, however, that a ${\mathbb Z}_{2}$ symmetry does not lead to a meaningful constraint. This is due to the Rohlin's theorem: for any smooth four-dimensional manifold that admits a spin structure ({\it i.e.}\/, its second Stiefel--Whitney class vanishes), its signature is divisible by 16. On the other hand, the Hirzebruch signature theorem says the signature of 4-dimensional manifolds is eight times the $\hat{A}$-genus which counts the number of fermion zero modes.  Therefore, the number of fermion zero modes is always even. It implies that the gravitational instanton leaves the fermion path integral invariant under any ${\mathbb Z}_{2}$ symmetries.

The Type-II anomalies match in all known well-established dualities, but it has a potential loophole. When the argument above is extended from non-abelian global symmetries to a $U(1)$ symmetry, we can still consider an instanton background, but the number of zero modes depends on the minimum unit of the $U(1)$ charges in the theory. In Ref.~\cite{Csaki:1997aw}, it was argued that we can take the smallest charge in the particle content of the theory. There is a caveat, however. In principle, some of the heavy states that are integrated out in the IR limit may have charge fractionalization. In this case the minimum unit of the $U(1)$ charges may become smaller, leading to weaker constraints.

Similarly, any abelian discrete symmetries can be embedded into $U(1)$s and we can consider triangle anomalies for $U(1)$ factors, and require that they match modulo the least common multiple among the relevant $N$'s. In addition, when the relevant $N$'s are all even, there is an important consideration. For ${\mathbb Z}_{N_{1}}={\mathbb Z}_{2k_{1}}$, ${\mathbb Z}_{N_{2}}={\mathbb Z}_{2k_{2}}$, and ${\mathbb Z}_{N_{3}}={\mathbb Z}_{2k_{3}}$, it is possible that a state transforms by $-1$ under all of the discrete symmetries (namely charge $(k_{1}, k_{2}, k_{3})$) so that it is allowed to have a Majorana mass and decouples, thereby shifting the anomaly by $k_{1}k_{2}k_{3} = \frac{1}{8} N_{1} N_{2} N_{3}$. In particular, it implies that there are no constraints considering ${\mathbb Z}_{2}^{3}$ anomalies because they can be shifted by 1 (see e.g. \cite{IBANEZ1991291, Banks:1991xj, Araki:2008ek}). This case is subject to the same caveat about charge fractionalization.

In the later sections, we consider both Type-I and Type-II anomalies.  They all match when they should, namely when the discrete symmetries are respected by the ground states.

\subsection{Seiberg Duality}

\begin{table}[t]
\renewcommand{\arraystretch}{1.6}
\setlength{\arrayrulewidth}{.2mm}
\setlength{\tabcolsep}{0.5em}
	\centerline{
	\begin{tabular}{|c|c|c|c|c|}
	\hline
	& $SU(N)$ & $U(1)_{R}$ & $SU(F)_{Q}$ & $SU(F)_{\tilde{Q}}$ \\ \hline
	$W_{\alpha}$ & {\bf adj} & $+1$ & ${\bf 1}$ & ${\bf 1}$ \\ \hline
	$Q$ & $\Yfund$ & $1-\frac{N}{F}$ & $\Yfund$ & ${\bf 1}$ \\ \hline
	$\tilde{Q}$ & $\overline{\Yfund}$ & $1-\frac{N}{F}$ & ${\bf 1}$ & $\Yfund$ \\ \hline
	\end{tabular}
	}
	\caption{Quantum numbers of fields in the electric $SU(N)$ SQCD.}\vspace{4mm}
    \label{tab:Seiberg1}
	\centerline{
	\begin{tabular}{|c|c|c|c|c|}
	\hline
	& $SU(F-N)$ & $U(1)_{R}$ & $SU(F)_{Q}$ & $SU(F)_{\tilde{Q}}$ \\ \hline
	$W_{\alpha}$ & {\bf adj} & $+1$ & ${\bf 1}$ & ${\bf 1}$ \\ \hline
	$q$ & $\Yfund$ & $1-\frac{\tilde{N}}{F}$ & $\overline{\Yfund}$ & ${\bf 1}$ \\ \hline
	$\tilde{q}$ & $\overline{\Yfund}$ & $1-\frac{\tilde{N}}{F}$ & ${\bf 1}$ & $\overline{\Yfund}$ \\ \hline
	$M=\tilde{Q}Q$ & ${\bf 1}$ & $2-\frac{2N}{F}$ & $\Yfund$ & $\Yfund$ \\ \hline
	\end{tabular}
	}
	\caption{Quantum numbers of fields in the magnetic $SU(\tilde{N})$ ($\tilde{N}=F-N$) SQCD.}
	\label{tab:Seiberg2}
\end{table}

The Seiberg duality in $SU(N)$ gauge theories~\cite{Seiberg:1994pq} states that electric and magnetic $SU(\tilde{N})$ ($\tilde{N}=F-N$) theories in \cref{tab:Seiberg1,tab:Seiberg2} are equivalent in the IR limit. The electric theory does not have a superpotential, while the magnetic theory has the superpotential
\begin{equation}
W = \frac{1}{\mu} M^{i j} \tilde{q}_{i} q_{j} \,.
\end{equation}
The meson field in the magnetic theory is matched to the composite in the electric theory as
\begin{equation}
M^{ij} = \tilde{Q}^{i} Q^{j} \,.
\end{equation}
It is highly non-trivial that the continuous symmetry anomalies $({\rm grav})^{2}U(1)_{R}$, $U(1)_{R}^{3}$, $U(1)_{R} SU(F)_{Q}^{2}$, $U(1)_{R} SU(F)_{\tilde{Q}}^{2}$ all match between the two theories.

In the following, we will investigate the discrete anomaly matching conditions associated with the charge conjugation $\CC_S$ and $\CC_A$. As explained in \cref{sec:twov,sec:para}, $\CC_S$ generally exists and is also non-anomalous. On the other hand, $\CC_A$ exists only for even $N$ and it is non-anomalous only for even $F$.

\subsubsection*{\boldmath $\CC_S SU(F)_C^2$ anomaly matching}

The charge conjugation $\CC_S$ interchanges $Q$ and $\tilde{Q}$ fields and hence does not commute with $SU(F)_Q$ or $SU(F)_{\tilde{Q}}$. Therefore, it is not clear how to study the anomalies such as $\CC_S SU(F)_Q^2$ or $\CC_S SU(F)_{\tilde{Q}}^2$. Instead, we study the anomaly $\CC_S SU(F)_C^2$, where $SU(F)_C$ is the diagonal subgroup $SU(F)_C \subset SU(F)_Q \times SU(F)_{\tilde{Q}}$ that commutes with $\CC_S$.\footnote{Note that the common convention is that $Q$ is in the fundamental representation under $SU(F)_Q$ and $\tilde{Q}$ is anti-fundamental under $SU(F)_{\tilde{Q}}$, but this is {\it not}\/ our convention here. For our convenience, we choose the definition of $SU(F)_{\tilde{Q}}$ such that $\tilde{Q}$ is fundamental under $SU(F)_{\tilde{Q}}$. In this way, $Q$ and $\tilde{Q}$ are on equal footing under the flavor symmetries; e.g. they are in the same representation under the diagonal subgroup $SU(F)_C$.}

Let us first look at this anomaly in the electric theory. The gaugino field $\lambda$ in $W_\alpha$ is a singlet under $SU(F)_C$, so it does not contribute. The quark and anti-quark fields $Q, \tilde{Q}$ both have $N$ zero modes under the instanton of $SU(F)_{C}$. Since $Q$ and $\tilde{Q}$ are interchanged, the linear combinations $Q\pm \tilde{Q}$ are even and odd eigenstates under $\CC_S$ respectively. Therefore, their contributions to the $\CC_S SU(F)_C^2$ anomaly is $N \tr(t^a t^b)=N\frac12\delta^{ab}$, where $t^a$ are generators of $SU(F)_C$ in the fundamental representation.

Similarly in the magnetic theory, the $\CC_S SU(F)_C^2$ anomaly receives a contribution $\tilde{N}\tr(t^a t^b)=\tilde{N}\frac12\delta^{ab}$ from magnetic quark and anti-quark fields $q, \tilde{q}$. In addition, there is a contribution from the meson field $M$, which transforms as $M\to UMU^T$ under elements in the diagonal subgroup $U\in SU(F)_C$. Clearly, $M$ decomposes into symmetric ($M^T=M$) and antisymmetric $M^T=-M$ representations. Under $\CC_S$, $Q$ and $\tilde{Q}$ fields are interchanged in the electric theory, and correspondingly in the magnetic theory the meson field $M$ is transposed. Obviously, the anomaly receives contributions only from the antisymmetric representation. The contribution is $\tr(t_\text{asym}^a t_\text{asym}^b)=(F-2)\frac12\delta^{ab}$.

Altogether, the anomalies do match between the electric and magnetic theories:
\begin{align}
N = \tilde{N} + (F-2) \quad \mbox{mod 2} \,.
\end{align}

\subsubsection*{\boldmath $\CC_S U(1)_R^2$ anomaly matching}

One can also study the $\CC_S U(1)_R^2$ anomalies. Since the $R$-charges are fractional, we use the $R$-charge normalized in the unit of $1/F$. The gaugino field $\lambda$ in $W_\alpha$ forms an adjoint representation of the gauge field $SU(N)$, which decomposes into $\Ysymm$ and $\Yasymm$ representations of its $\CC_S$-invariant subgroup $SO(N)$. Since the gaugino field transforms as $\lambda \rightarrow - \lambda^{T}$ under charge conjugation (see e.g. \cref{eqn:Atransform}), it is the $\Ysymm$ component that is odd under $\CC_S$. Their contributions to the $\CC_S U(1)_R^2$ anomaly are therefore $\left[\frac12 N(N+1)-1\right]F^2$ and $\left[\frac12 \tilde{N}(\tilde{N}+1)-1\right]F^2$ in the electric and magnetic theories respectively.

For the quark and anti-quark fields, note that the $R$-charge listed in \cref{tab:Seiberg1,tab:Seiberg2} are those for the chiral superfields; the $R$-charge for the fermionic components are obtained by further adding $-1$ on top of them. After normalizing the $R$-charge to the unit of $1/F$, we obtain their contributions to the $\CC_S U(1)_R^2$ anomaly as $FN\times N^2$ and $F\tilde{N}\times \tilde{N}^2$ in the electric and magnetic theories respectively.

Finally, we also have a contribution $\frac12 F(F-1)(F-2N)^2$ from the meson field $M$ in the magnetic theory. Putting everything together, one can check that the anomaly does match between the electric and magnetic theories:
\begin{align}
&\left[\frac{1}{2}N(N+1)-1\right] F^2 + F N \times N^2 \notag\\[3pt]
&= \left[\frac12\tilde{N}(\tilde{N}+1)-1\right] F^2 + F\tilde{N}\times\tilde{N}^2 + \frac12F(F-1) (F-2N)^2 \quad\mbox{ mod 2} \,.
\end{align}

\subsubsection*{\boldmath Anomaly matching associated with $\CC_A$}

When $N$ and $F$ are both even, we can also consider anomalies associated with the charge conjugation $\CC_A$. However, since all integers involved will be even, the matching conditions will work out trivially. This fact is consistent with the observation that $\CC_A$ is equivalent to $\CC_S$ up to the sign flips of all $\tilde{Q}$ which is non-anomalous when $F$ is even.

\subsection{Kutasov Duality}

\begin{table}[t]
\renewcommand{\arraystretch}{1.6}
\setlength{\arrayrulewidth}{.2mm}
\setlength{\tabcolsep}{0.5em}
	\centerline{
	\begin{tabular}{|c|c|c|c|c|}
	\hline
	& $SU(N)$ & $U(1)_{R}$ & $SU(F)_{Q}$ & $SU(F)_{\tilde{Q}}$ \\ \hline
	$W_{\alpha}$ & {\bf adj} & $+1$ & ${\bf 1}$ & ${\bf 1}$ \\ \hline
	$X$ & {\bf adj} & $\frac{2}{k+1}$ & ${\bf 1}$ & ${\bf 1}$ \\ \hline
	$Q$ & $\Yfund$ & $1- \frac{2N}{(k+1)F}$ & $\Yfund$ & ${\bf 1}$ \\ \hline
	$\tilde{Q}$ & $\overline{\Yfund}$ & $1- \frac{2N}{(k+1)F}$ & ${\bf 1}$ & $\Yfund$ \\ \hline
	\end{tabular}
	}
	\caption{Quantum numbers of fields in the electric $SU(N)$ SQCD with an adjoint.}\vspace{4mm}
    \label{tab:Kutasov1}
	\centerline{
	\begin{tabular}{|c|c|c|c|c|}
	\hline
	& $SU(kF-N)$ & $U(1)_{R}$ & $SU(F)_{Q}$ & $SU(F)_{\tilde{Q}}$ \\ \hline
	$W_{\alpha}$ & {\bf adj} & $+1$ & ${\bf 1}$ & ${\bf 1}$ \\ \hline
	$Y$ & {\bf adj} & $\frac{2}{k+1}$ & ${\bf 1}$ & ${\bf 1}$ \\ \hline
	$q$ & $\Yfund$ & $1- \frac{2\tilde{N}}{(k+1)F}$ & $\overline{\Yfund}$ & ${\bf 1}$ \\ \hline
	$\tilde{q}$ & $\overline{\Yfund}$ & $1- \frac{2\tilde{N}}{(k+1)F}$ & ${\bf 1}$ & $\overline{\Yfund}$
	\\ \hline
	$M_{j}=\tilde{Q}X^{j}Q$ & ${\bf 1}$ & $2+\frac{2jF-4N}{(k+1)F}$ & $\Yfund$ & $\Yfund$ \\ \hline
	\end{tabular}
	}
	\caption{Quantum numbers of fields in the magnetic $SU(\tilde{N})$ ($\tilde{N}=kF-N$) SQCD with an adjoint.}
	\label{tab:Kutasov2}
\end{table}

The Kutasov duality~\cite{Kutasov:1995ve} is a generalization of Seiberg duality when an adjoint superfield $X$ is added with the superpotential
\begin{align}
W = \Tr \left(X^{k+1}\right) \,.
\end{align}
The dual magnetic theory is based on the gauge group $SU(\tilde{N})$ ($\tilde{N} = k F-N$) with the adjoint superfield $Y$ and meson fields $M_j = \tilde{Q} X^j Q$ for $j=0, \cdots, k-1$ and the superpotential
\begin{align}
W = \Tr \left(Y^{k+1}\right) + \sum_{j=0}^{k-1} M_j\, \tilde{q}\, Y^{k-1-j}\, q \,.
\end{align}
The field representations in the electric and magnetic theories in the Kutasov duality are summarized in \cref{tab:Kutasov1,tab:Kutasov2}. In order for the superpotential to be invariant for all $k$, we assign the odd intrinsic charge conjugation to $X$ so that $X \rightarrow + X^{T}$, and similarly for $Y$. Note that in the limit $k\rightarrow 1$, the Kutasov duality reduces to the Seiberg duality, because the superpotential $\Tr \left(X^2\right)$ and $\Tr \left(Y^2\right)$ are mass terms for the new adjoint fields, which can be integrated out and decouple from the rest of the theory.

\subsubsection*{\boldmath $\CC_S SU(F)_C^2$ anomaly matching}

Compared to the case of Seiberg duality, the newly added adjoint fields $X$ and $Y$ are singlets under the flavor symmetries and hence do not contribute to the $\CC_S SU(F)_C^2$ anomaly. Contributions from other fields are essentially the same as we discussed in Seiberg duality, except that we now have a different value of $\tilde{N}=kF-N$, and that we have more meson fields in the dual magnetic theory. The anomaly matching condition does work out:
\begin{align}
N = (kF-N) + \sum_{j=0}^{k-1} (F-2) \quad \mbox{mod 2} \,.
\end{align}

\subsubsection*{\boldmath $\CC_S U(1)_R^2$ anomaly matching}

The $R$-charges now come in the unit of $\frac{1}{(k+1)F}$. One can check that the $\CC_S U(1)_R^2$ anomalies are also matched:
\begin{align}
&\left[\frac12 N(N+1)-1\right] \left[(k+1)F\right]^2 + \left[\frac12 N(N-1)\right] \left[2F-(k+1)F\right]^2 + FN(2N)^2 \notag\\[5pt]
&= \left[\frac12\tilde{N}(\tilde{N}+1)-1\right] \left[(k+1)F\right]^2 + \left[\frac12\tilde{N}(\tilde{N}-1)\right] \left[2F-(k+1)F\right]^2 + F\tilde{N}\left[2\tilde{N}\right]^2 \notag\\[3pt]
&\quad + \frac12 F(F-1) \sum_{j=0}^{k-1} \left[(k+1+2j)F-4N\right]^2 \quad\mbox{ mod 2} \,.
\end{align}

\subsection{An example with anti-symmetric tensor}

\begin{table}[t]
\renewcommand{\arraystretch}{1.6}
\setlength{\arrayrulewidth}{.2mm}
\setlength{\tabcolsep}{0.5em}
	\centerline{
	\begin{tabular}{|c|c|c|c|c|c|c|}
	\hline
	& $SU(N)$ & $SU(F)_{Q}$ & $SU(F)_{\tilde{Q}}$ & $U(1)_{X}$ & $U(1)_{B}$ & $U(1)_{R}$
	\\ \hline
	$W_{\alpha}$ & {\bf adj} & ${\bf 1}$ & ${\bf 1}$ & $0$ & $0$ & $+1$\\ \hline
	$X$ & $\Yasymm$ & ${\bf 1}$ & ${\bf 1}$ & $1$ & $\frac{2}{N}$ & $\frac{1}{k+1}$\\ \hline
	$\tilde{X}$ & $\overline{\Yasymm}$ & ${\bf 1}$ & ${\bf 1}$ & $-1$ & $-\frac{2}{N}$
	& $\frac{1}{k+1}$\\ \hline
	$Q$ & $\Yfund$ & $\Yfund$ & {\bf 1} & $0$ & $\frac{1}{N}$ &
	$1-\frac{N+2k}{(k+1)F}$ \\ \hline
	$\tilde{Q}$ & $\overline{\Yfund}$ & {\bf 1} & $\Yfund$ & $0$ & $-\frac{1}{N}$ &
	$1-\frac{N+2k}{(k+1)F}$ \\ \hline
	\end{tabular}
	}
	\caption{Quantum numbers of fields in the electric $SU(N)$ SQCD with rank-two anti-symmetric tensors $X$ and $\tilde{X}$.}\vspace{4mm}
    \label{tab:ILS1}
	\centerline{
	\begin{tabular}{|c|c|c|c|c|c|c|}
	\hline
	& $SU(\tilde{N})$ & $SU(F)_{Q}$ & $SU(F)_{\tilde{Q}}$
	& $U(1)_{X}$ & $U(1)_{B}$ & $U(1)_{R}$
	\\ \hline
	$W_{\alpha}$ & {\bf adj} & ${\bf 1}$ & ${\bf 1}$ & $0$ & $0$ & $+1$\\ \hline
	$Y$ & $\Yasymm$ & ${\bf 1}$ & ${\bf 1}$ & $\frac{N-F}{\tilde{N}}$ & $\frac{2}{\tilde{N}}$
	& $\frac{1}{k+1}$\\ \hline
	$\tilde{Y}$ & $\overline{\Yasymm}$ & ${\bf 1}$ & ${\bf 1}$ & $-\frac{N-F}{\tilde{N}}$
	& $-\frac{2}{\tilde{N}}$ & $\frac{1}{k+1}$\\ \hline
	$q$ & $\Yfund$ & $\overline{\Yfund}$ & ${\bf 1}$ & $\frac{k(F-2)}{\tilde{N}}$
	& $\frac{1}{N}$ & $1-\frac{\tilde{N}+2k}{(k+1)F}$ \\ \hline
	$\tilde{q}$ & $\overline{\Yfund}$ & ${\bf 1}$ & $\overline{\Yfund}$
	& $-\frac{k(F-2)}{\tilde{N}}$ & $-\frac{1}{N}$ & $1-\frac{\tilde{N}+2k}{(k+1)F}$ \\ \hline
	$M_{j} = \tilde{Q} (\tilde{X}X)^{j} Q$ & ${\bf 1}$ & $\Yfund$ & $\Yfund$ & $0$
	& $0$ & $\frac{\tilde{N}-N+(2j+1)F}{(k+1)F}$ \\ \hline
	$P_{r} = Q (\tilde{X}X)^{r} \tilde{X} Q$ & ${\bf 1}$ & $\Yasymm$ & ${\bf 1}$ & $-1$
	& $0$ & $\frac{\tilde{N}-N+2(r+1)F}{(k+1)F}$ \\ \hline
	$\tilde{P}_{r} = \tilde{Q} (\tilde{X}X)^{r} X \tilde{Q}$ & ${\bf 1}$ & ${\bf 1}$ & $\Yasymm$ & $1$
	& $0$ & $\frac{\tilde{N}-N+2(r+1)F}{(k+1)F}$ \\ \hline
	\end{tabular}
	}
	\caption{Quantum numbers of fields in the magnetic $SU(\tilde{N})$ ($\tilde{N}=(2k+1)F-4k-N$) SQCD with rank-two anti-symmetric tensors $Y$ and $\tilde{Y}$.}
	\label{tab:ILS2}
\end{table}

A non-trivial example of duality with an anti-symmetric tensor was found by Intriligator, Leigh, and Strassler in \cite{Intriligator:1995ax}. The electric gauge group is $SU(N)$ with the superpotential
\begin{align}
W = \Tr \left[(X \tilde{X})^{k+1}\right] \,,
\end{align}
while the magnetic theory has the gauge group $SU(\tilde{N})$ ($\tilde{N}=(2k+1)F-4k-N$) and the superpotential
\begin{align}
W &= \Tr\left[(Y \tilde{Y})^{k+1}\right] + \sum_{j=0}^{k} M_{k-j}\, q\, (\tilde{Y} Y)^j\, \tilde{q} \notag\\
&\quad + \sum_{r=0}^{k-1} \left[ P_{k-1-r}\, q\, (\tilde{Y}Y)^r\, \tilde{Y}\, q + \tilde{P}_{k-1-r}\, \tilde{q}\, (Y\tilde{Y})^r\, Y\, \tilde{q} \right] \,.
\end{align}
The field representations are summarized in \cref{tab:ILS1,tab:ILS2}. Under charge conjugation, the rank-two anti-symmetric tensor $X$ and its conjugate representation $\tilde{X}$ are exchanged, so are $Y$ and $\tilde Y$; the fields $P_r$ and $\tilde{P}_r$ also get interchanged; transformations of other fields are the same as in the Seiberg duality. When $k=0$, this example reduces to Seiberg duality.

For discrete anomaly matching conditions, $U(1)_{X}$ and $U(1)_{B}$ do not commute with $\CC_S$. The $\CC_S SU(F)_C^2$ anomalies match as
\begin{align}
N = \tilde{N} + \sum_{j=0}^{k} (F-2) + \sum_{r=0}^{k-1} (F-2) \quad\mbox{ mod 2} \,.
\end{align}
The $\CC_S U(1)_R^2$ anomalies match as
\begin{align}
&\left[\frac12 N(N+1)-1\right] \left[(k+1)F\right]^2 + \left[\frac12 N(N-1)\right] \left[F-(k+1)F\right]^2 + FN(N+2k)^2 \notag\\[8pt]
&= \left[\frac12 \tilde{N}(\tilde{N}+1)-1\right] \left[(k+1)F\right]^2 + \left[\frac12\tilde{N}(\tilde{N}-1)\right] \left[F-(k+1)F\right]^2 + F\tilde{N}(\tilde{N}+2k)^2 \notag\\[3pt]
&\quad + \frac12 F(F-1) \sum_{j=0}^k \left[\tilde{N}-N+(2j+1)F-(k+1)F\right]^2 \notag\\
&\quad + \frac12 F(F-1) \sum_{r=0}^{k-1} \left[\tilde{N}-N+2(r+1)F-(k+1)F\right]^2 \quad\mbox{ mod 2} \,.
\end{align}

\subsection{Parity in $SO(2r)$ gauge theories}

\begin{table}[t]
\renewcommand{\arraystretch}{1.5}
\setlength{\arrayrulewidth}{.2mm}
\setlength{\tabcolsep}{0.5em}
	\centerline{
	\begin{tabular}{|c|c|c|c|c|}
	\hline
	& $SO(N)$ & $U(1)_{R}$ & $SU(F)_{Q}$ & ${\mathbb Z}_{2F}$ \\ \hline
	$W_{\alpha}$ & {\bf adj} & $+1$ & ${\bf 1}$ & $0$ \\ \hline
	$Q$ & $\Yfund$ & $1-\frac{N-2}{F}$ & $\Yfund$ & $1$ \\ \hline
	\end{tabular}
	}
	\caption{Quantum numbers of fields in the electric $SO(N)$ SQCD.}\vspace{4mm}
    \label{tab:Intriligator-Seiberg1}
	\centerline{
	\begin{tabular}{|c|c|c|c|c|}
	\hline
	& $SO(\tilde{N})$ & $U(1)_{R}$ & $SU(F)_{Q}$ & ${\mathbb Z}_{2F}$ \\ \hline
	$W_{\alpha}$ & {\bf adj} & $+1$ & ${\bf 1}$ & $0$ \\ \hline
	$q$ & $\Yfund$ & $1-\frac{\tilde{N}-2}{F}$ & $\overline{\Yfund}$ & $-1$ \\ \hline
	$M=QQ$ & ${\bf 1}$ & $2-2\frac{N-2}{F}$ & $\Ysymm$ & $+2$ \\ \hline
	\end{tabular}
	}
	\caption{Quantum numbers of fields in the magnetic $SO(\tilde{N})$ ($\tilde{N}=F-N+4$) SQCD.}
	\label{tab:Intriligator-Seiberg2}
\end{table}

It is instructive to study anomalies associated with the parity outer automorphism in $SO(2r)$ gauge theories.   The electric-magnetic duality found by Intriligator and Seiberg \cite{Intriligator:1995id} says ${\cal N}=1$ supersymmetric $SO(N)$ gauge theory with $F$ chiral superfields in the vector representation is dual to $SO(F-N+4)$ gauge theory also with $F$ vectors and a meson superfield $M$ in the rank-two symmetric representation of $SU(F)$.  See \cref{tab:Intriligator-Seiberg1,tab:Intriligator-Seiberg2} for the particle contents.

One special aspect is the ${\mathbb Z}_{2F}$ discrete symmetry.  The matching of baryon operators between the electric and magnetic theories is given by
\begin{align}
B_{\ell_{1} \cdots \ell_{N-4}} &= \frac{1}{N!}\, \epsilon_{i_{1}\cdots i_{N}}\, Q^{i_{1}}_{\ell_{1}} \cdots Q^{i_{N-4}}_{\ell_{N-4}}\, W^{i_{N-3}i_{N-2}}_{\alpha}\, W^{\alpha i_{N-1}i_{N}} \notag\\[5pt]
&= \frac{1}{(F-N+4)!}\, \frac{1}{F!}\, \epsilon^{i_{1}\cdots i_{F-N+4}}\, \epsilon_{\ell_{1} \cdots \ell_{N-4} \ell_{N-3} \cdots \ell_{F}}^{}\, q_{i_{1}}^{\ell_{N-3}} \cdots q_{i_{F-N+4}}^{\ell_{F}} \,.
\end{align}
Here, $i_{k}$ are the gauge indices while $\ell_{k}$ are the flavor indices.  Its ${\mathbb Z}_{2F}$ charge in the electric theory is $N-4$, while that in the magnetic theory is $-\tilde{N} = -(F-N+4) = N-4-F$.  The mismatch is fixed by mapping the ${\mathbb Z}_{2F}$ in the electric theory to ${\mathbb Z}_{2F} {\cal P}$ in the magnetic theory, where ${\cal P}$ provides a minus sign for the Levi-Civita symbol for the gauge indices which can be identified as charge $F$ under the ${\mathbb Z}_{2F}$ symmetry.  It was verified in Ref.~\cite{Csaki:1997aw} that all Type-I and Type-II anomalies match under this identification.

The parity ${\cal P}$ is an independent symmetry only when both $N$ and $F$ are even.  When $N$ is odd, we can combine ${\cal P}$ with the rotation ${\rm diag}(-1,-1, \cdots, -1, -1, +1)$ in $SO(N)$, which becomes an overall sign change of $Q$.  It is then a subgroup of ${\mathbb Z}_{2F}$, not a new symmetry.  When $F$ is odd and $N$ even, ${\mathbb Z}_{2F} = {\mathbb Z}_{F} \times {\mathbb Z}_{2}$ since 2 and $F$ are relatively prime, and ${\mathbb Z}_{F}$ is the center group of $SU(F)$, while ${\mathbb Z}_{2}$ is a part of the center group of $SO(N)$.

Now we can verify anomaly matching conditions under ${\cal P}$. Only one of the components of the vector representation switches its sign. The ${\cal P} SU(F)^{2}$ anomalies are matched as
\begin{align}
1 = 1 \quad\mbox{ mod 2} \,.
\end{align}
The ${\cal P} {\mathbb Z}_{2F}^{2}$ anomalies are matched as
\begin{align}
F \times 1^2 = F \times (-1)^2 \quad\mbox{ mod 2} \,.
\end{align}
The ${\cal P} U(1)_{R}^{2}$ anomalies are matched, on the other hand, using the $1/F$ as the minimum charge as
\begin{align}
(N-1) F^2 + F(N-2)^2 = (\tilde{N}-1) F^2 + F (\tilde{N}-2)^2 \quad\mbox{ mod 2} \,.
\end{align}
Here, the first terms on both sides of the equation are the $N-1$ ($\tilde{N}-1$) vector of the $\PP$-invariant $SO(N-1)$ ($SO(\tilde{N}-1)$) subgroup of gauginos in the adjoint representation, which flip sign under the parity ${\cal P}$.

\subsection{$SO(8)$ s-confining theory}

\begin{table}[t]
\renewcommand{\arraystretch}{1.5}
\setlength{\arrayrulewidth}{.2mm}
\setlength{\tabcolsep}{0.5em}
	\centerline{
	\begin{tabular}{|c|c|c|c|c|c|c|}
	\hline
	& $SO(8)$ & $SU(3)$ & $SU(3)$ & $U(1)_{1}$ & $U(1)_{2}$ & $U(1)_{R}$\\ \hline
	$W_{\alpha}$ & {\bf adj} & ${\bf 1}$ & ${\bf 1}$ & $0$ & $0$ & $+1$\\ \hline
	$Q$ & $8_{v}$ & ${\bf 1}$ & ${\bf 1}$ & $0$ & $6$ & $1$ \\ \hline
	$S$ & $8_{s}$ & ${\bf 3}$ & ${\bf 1}$ & $1$ & $-1$ & $0$ \\ \hline
	$S'$ & $8_{c}$ & ${\bf 1}$ & ${\bf 3}$ & $-1$ & $-1$ & $0$ \\ \hline \hline
	$Q^{2}$ & & ${\bf 1}$ & ${\bf 1}$ & $0$ & $12$ & $2$ \\ \hline
	$S^{2}$ & & $\Ysymm$ & ${\bf 1}$ & $2$ & $-2$ & $0$ \\ \hline
	$S^{\prime 2}$ & & ${\bf 1}$ & $\Ysymm$ & $-2$ & $-2$ & $0$ \\ \hline
	$S S' Q$ & & $\Yfund$ & $\Yfund$ & $0$ & $4$ & $1$ \\ \hline
	$S^{3} S' Q$ & & ${\bf 1}$ & $\Yfund$ & $2$ & $2$ & $1$ \\ \hline
	$S^{\prime 3} S Q$ & & $\Yfund$ & ${\bf 1}$ & $-2$ & $2$ & $1$ \\ \hline
	$S^{2} S^{\prime 2}$ & & $\overline{\Yfund}$ & $\overline{\Yfund}$ & $0$ & $-4$ & $0$
	\\ \hline
	\end{tabular}
	}
	\caption{Quantum numbers of fields in the s-confining $SO(8)$ theory and massless composites.}
	\label{tab:CSW}
\end{table}

Here is another non-trivial example where we can test the parity ${\cal P}$.  It is an $SO(8)$ theory with three spinors, three conjugate spinors, and one vector \cite{Csaki:1996zb}. Under parity, the $\PP$-invariant subgroup is $SO(7)$. See \cref{tab:CSW} for the particle content and quantum numbers. $S$ and $S'$ are interchanged, while the gauge multiplet decomposes as ${\bf adj}_8= {\bf adj}_7 + {\bf 7}$ and the ${\bf 7}$ changes its sign.  Two $SU(3)$ are interchanged, and we consider the diagonal subgroup $SU(3)_{C}$. $U(1)_1$ does not commute with parity and we do not consider its anomaly.

\noindent\textbf{\boldmath ${\cal P}SU(3)_{C}^{2}$:}  Note that $SS'Q$ decomposes as $\Ysymm + \Yasymm$ under $SU(3)_{C}$ and $\Yasymm$ is odd under parity. Similarly for $S^{2} S^{\prime 2}$.  The linear combinations $S-S'$, $S^{2} - S^{\prime 2}$, $S^{3} S' Q - S^{\prime 3} S Q$ are odd. The matching condition works out as
\begin{align}
8_{S-S'} &= 5_{S^2-S^{\prime 2}} + 1_{S S' Q(\Yasymm)} + 1_{S^3 S' Q - S^{\prime 3} S Q} + 1_{S^2 S^{\prime 2} (\Yfund)} \quad\mbox{ mod 2} \,.
\end{align}

\noindent\textbf{\boldmath ${\cal P}U(1)_{2}^{2}$:} Note that one component of $Q$ changes its sign under parity. The matching condition works out as
\begin{align}
&1 \times 6^2_Q + 8 \times 3 (-1)^2_{S-S'} \notag\\[8pt]
&= 6 (-2)^2_{S^2-S^{\prime 2}} + 3 \times 4^2_{S S' Q(\Yasymm)} + 3 \times 2^2_{S^3 S' Q - S^{\prime 3} S Q} + 3 (-4)^2_{S^2 S^{\prime 2} (\Yfund)} \quad\mbox{ mod 2} \,.
\end{align}

\noindent\textbf{\boldmath ${\cal P}U(1)_{R}^{2}$:} Here, gauginos also contribute. The anomalies do match
\begin{align}
7 \times 1^2_\lambda + 8 \times 3 (-1)^2_{S-S'} = 6 (-1)^2_{S^2-S^{\prime 2}} + 3 \times (-1)^2_{S^2 S^{\prime 2} (\Yfund)} \quad\mbox{ mod 2} \,.
\end{align}

\noindent\textbf{\boldmath ${\cal P}U(1)_{2} U(1)_{R}$:} This anomaly matching also works out
\begin{align}
8 \times 3 (-1)(-1)_{S-S'} = 6 (-2)(-1)_{S^2-S^{\prime 2}} + 3 (-4)(-1)_{S^2 S^{\prime 2} (\Yfund)} \quad\mbox{ mod 2} \,.
\end{align}

\section{Spontaneous Breaking of Outer Automorphism}
\label{sec:spont}

Here we discuss two theories that exhibit similar dynamics where the theory confines with a moduli space of vacua.  Yet in one example the charge conjugation is spontaneously broken, while in the other it is not.  Both of them are ${\cal N}=1$ supersymmetric gauge theories.

\subsection{$SO(6)$ with two vectors}

\begin{table}[t]
\renewcommand{\arraystretch}{1.5}
\setlength{\arrayrulewidth}{.2mm}
\setlength{\tabcolsep}{0.5em}
	\centerline{
	\begin{tabular}{|c|c|c|c|c|}
	\hline
	& $SO(6)$ & $U(1)_{R}$ & $SU(2)_{f}$ & ${\mathbb Z}_{4}$ \\ \hline \hline
	$W_{\alpha}$ & {\bf adj} & $+1$ & ${\bf 1}$ & 0 \\ \hline
	$\phi_{i}$ & ${\bf 6}$ & $-1$ & ${\bf 2}$ & 1 \\ \hline \hline
	$M_{ij}$ & {\bf 1} & $-2$ & ${\bf 3}$ & 2 \\ \hline
	$W_{\alpha}W^{\alpha}$ & {\bf 1} & $+2$ & ${\bf 1}$ & 0 \\ \hline
	${\cal O}$ & {\bf 1} & $0$ & ${\bf 1}$ & 2 \\ \hline
	\end{tabular}
	}
	\caption{Quantum numbers of various fields in the $SO(6)$ theory with two vectors $\phi_{i}$.  The first two fields are in the UV theory, while the last three are in the IR theory.  The last row is the operator ${\cal O}=\epsilon_{abcdef} \epsilon^{\alpha\beta}W_{\alpha}^{ab} W_{\beta}^{bc} \phi_{i}^{e} \phi_{j}^{f} \epsilon^{ij}$ that acquires an expectation value and breaks ${\mathbb Z}_{4}$ to ${\mathbb Z}_{2}$ \cite{Csaki:1997aw}.}\label{tab:SO62vec1}\vspace{4mm}
	\centerline{
	\begin{tabular}{|c|c|c|c|c|c|c|c|}
	\hline
	& $R$(grav)$^{2}$ & $R^{3}$ & $R(SU(2))^{2}$ & ${\mathbb Z}_{4}(SU(2))^{2}$
	& ${\mathbb Z}_{4} R^{2}$ & ${\mathbb Z}_{4}^{2} R$ & ${\mathbb Z}_{4}^{3}$
	\\ \hline \hline
	$W_{\alpha}$ & $+15$ & $+15$ & 0 & 0 & 0 & 0 & 0 \\ \hline
	$\phi_{i}$ & $-24$ & $-96$ & $-6$ & 6 & 48 & $-24$ & 12 \\ \hline
	UV total & $-9$ & $-81$ & $-6$ & 6 & 48 & $-24$ & 12 \\ \hline \hline
	$M_{ij}$ & $-9$ & $-81$ & $-6$ & 4 & 54 & $-18$ & 24 \\ \hline
	\end{tabular}
	}
	\caption{Anomalies between the UV and IR particle contents.  The anomalies of continuous symmetries all match \cite{Intriligator:1995id}.  The ${\mathbb Z}_{4}$ anomalies are supposed to be matched mod 4, but they match only mod 2, indicating the spontaneous breaking due to $\langle {\cal O} \rangle$.}\label{tab:SO62vec2}\vspace{4mm}
	\centerline{
	\begin{tabular}{|c|c|c|c|}
	\hline
	& $SO(5)$ & $\CC_{A} R^{2}$ & $\CC_{A} (SU(2))^{2}$
	\\ \hline \hline
	$W_{\alpha}$ & ${\bf 10}_{+}\oplus{\bf 5}_{-}$ & $-$ & $+$   \\ \hline
	$\phi_{i}$ & ${\bf 5}_{+}\oplus{\bf 1}_{-}$ & $+$ & $-$   \\ \hline
	UV total & & $-$ & $-$ \\ \hline \hline
	$M_{ij}$ & ${\bf 1}_{+}$ & $+$ & $+$  \\ \hline
	\end{tabular}
	}
	\caption{The first column is the quantum number under the $\PP$-invariant subgroup $Sp(4)=SO(5)$. Anomalies of charge conjugation do not match, indicating that the charge conjugation is spontaneously broken.  These anomalies were not studied before.}
\label{tab:SO62vec3}
\end{table}

The first example is $SO(6)$ with two vectors $\phi_{i}$, $i=1,2$. This is a well-known example by Intriligator and Seiberg. The field representations are summarized in \cref{tab:SO62vec1}. Once two vectors go along the $D$-flat direction, $SO(6)$ generically breaks to pure $SO(4) \simeq SU(2)_{1} \times SU(2)_{2}$ Yang-Mills.  The matching condition is that the $SU(2)$ dynamical scale is given by the $SO(6)$ dynamical scale as
\begin{align}
\Lambda_{2}^{6} = \frac{\Lambda_{6}^{10}}{{\rm det} M} \,, \qquad
M_{ij} = \phi_{i} \phi_{j} \,.
\end{align}
The point here is that the outer automorphism of $SO(6)$, namely parity, is equivalent to interchange of two $SU(2)$'s, which leaves the $\PP$-invariant subgroup $SO(5)$, under which the $\phi_{i}$ decomposes into ${\bf 5}\oplus {\bf 1}$ as expected under parity. Two $SU(2)$ factors develop gaugino condensates and hence the superpotential is
\begin{align}
W = \pm \Lambda_{2,1}^{3} \pm \Lambda_{2,2}^{3} = (\pm 1 \pm 1) \frac{\Lambda_{6}^{5}}{({\rm det} M)^{1/2}} \,.
\end{align}
Note that the signs are coming from the square root of the dynamical scale $\Lambda_{2}^{6}$ above, and not correlated, as each $SU(2)$ factors lead to two vacua and hence there are four configurations altogether.  When two signs are the same, namely $(+,+)$ or $(-,-)$, the superpotential has run-away behavior for the meson superfields and there is no ground state; similar to the Affleck--Dine--Seiberg superpotential in $SU(2)$ theory with one flavor.  In this case, the charge conjugation that interchanges two $SU(2)$ factors is unbroken. These two run-away directions correspond to ${\mathbb Z}_{4}/{\mathbb Z}_{2}$ while the charge conjugation is unbroken.

When two signs are the opposite, namely $(+,-)$ or $(-,+)$, the meson dependence cancels exactly in the superpotential, and we have a moduli space of vacua.  In this case, the charge conjugation is spontaneously broken and hence there are two ground states; one can even have domain walls (probably BPS). The order parameter of charge conjugation (or equivalent parity) breaking is
\begin{align}
\langle \epsilon_{abcdef} \epsilon^{\alpha\beta}W_{\alpha}^{ab} W_{\beta}^{bc} \phi_{i}^{e} \phi_{j}^{f} \epsilon^{ij}
\rangle \neq 0 \,,
\end{align}
where $a,b,c,d,e,f=1,\cdots, 6$ are the $SO(6)$ indices, as shown in \cite{Csaki:1997aw}.\footnote{This condensate also breaks the ${\mathbb Z}_{4}$ discrete symmetry as reflected in \cref{tab:SO62vec2}.} In fact, the anomaly matching for the charge conjugation fails in this case, as shown in \cref{tab:SO62vec3}. These two vacua correspond to ${\cal C} \times {\mathbb Z}_{4}/{\mathbb Z}_{4}$ where the generator of the unbroken ${\mathbb Z}_4$ is embedded as $(-, \omega) \in {\cal C} \times {\mathbb Z}_4$ with $\omega = i$.

\subsection{$SU(6)$ with one rank-three anti-symmetric tensor}

\begin{table}[t]
\renewcommand{\arraystretch}{1.8}
\setlength{\arrayrulewidth}{.2mm}
\setlength{\tabcolsep}{0.5em}
	\centerline{
	\begin{tabular}{|c|c|c|c|c|}
	\hline
	& $SU(6)$ & $U(1)_{R}$ & ${\mathbb Z}_{6}$ & $SO(6)$ \\ \hline \hline
	$W_{\alpha}$ & {\bf adj} & $+1$ & 0 & $\Ysymm_{-}\oplus \Yasymm_{+}$ \\ \hline
	$A$ & $\Ythreea$ & $-1$ & 1 & $\Ythreea_{+} \oplus \Ythreea_{-}$ \\ \hline \hline
	$A^{4}$ & ${\bf 1}$ & $-4$ & 4 & ${\bf 1}_{+}$ \\ \hline
	$W_{\alpha} W^{\alpha}$ & ${\bf 1}$ & $+2$ & 0 & ${\bf 1}_{+}$ \\ \hline
	${\cal O}$ & {\bf 1} & 0 & 2 & ${\bf 1}_{+}$ \\ \hline
	\end{tabular}
	}
\caption{Quantum number of fields in the $SU(6)$ theory with rank-three anti-symmetric tensor $A$.  The last row is the operator ${\cal O} = \Tr \left(T^{a} A A\right) \Tr\left(T^{a} W_{\alpha}W^{\alpha}\right)$ that acquires an expectation value and breaks ${\mathbb Z}_{6}$ to ${\mathbb Z}_{2}$.}\label{tab:SU6rank31}\vspace{4mm}
	\centerline{
	\begin{tabular}{|c|c|c|c|c|c|c|}
	\hline
	& $R$(grav)$^{2}$ & $R^{3}$ & ${\mathbb Z}_{6}$(grav)$^{2}$
	& ${\mathbb Z}_{6} R^{2}$ & ${\mathbb Z}_{6}^{2} R$
	& $\CC_{S} R^{2}$ \\ \hline \hline
	$W_{\alpha}$ & 35 & 35 & 0 & 0 & 0 & $+$ \\ \hline
	$A$ & $-40$ & $-160$ & 40 & 80 & $-40$ & $+$  \\ \hline
	UV total & $-5$ & $-125$ & 4 & 2 & 4 & $+$ \\ \hline \hline
	$A^{4}$ & $-5$ & $-125$ & 8 & $100$ & $-80$ & $+$ \\ \hline
	\end{tabular}
	}
\caption{Anomaly matching conditions between the UV and IR.  The $R$-charge of $A$ is $-1$ so that $U(1)_{R}$ is anomaly free under the $SU(6)$ gauge group.  Note that the ${\mathbb Z}_{6}$ anomalies are not matched because it is broken to ${\mathbb Z}_{2}$ due to the $A^{2} W W$ condensate.  On the other hand, the charge conjugation anomalies had not been studied before, and are matched correctly.}
\label{tab:SU6rank32}
\end{table}

The second example is $SU(6)$ with one rank-three anti-symmetric tensor $A^{ijk}$. The field representations are summarized in \cref{tab:SU6rank31}. Since this is a pseudo-real representation, there is no $A^{2}$ invariant; the lowest invariant is $A^{4}$.  Along the $D$-flat direction, it breaks $SU(6)$ to $SU(3)_{1} \times SU(3)_{2}$ with the matching condition
\begin{align}
\Lambda_{3}^{9} = \pm \frac{\Lambda_{6}^{15}}{(A^{4})^{3/2}} \,.
\label{eq:sign}
\end{align}
Note the unusual square root in the denominator that leads to the sign ambiguity.  We show below it is not an ambiguity; two low-energy $SU(3)$ groups have opposite signs in their dynamical scales.

The representation $A^{ijk}$ has Dynkin index 6. Under $\CC_A$, it decomposes into one rank-three tensor (Dynkin index 5) and one fundamental representation (Dynkin index 1) under the $\CC_A$-invariant subgroup $Sp(6)$.  Under either assignment of even and odd eigenvalues, there is an odd number of zero modes under an $Sp(6)$ instanton, and hence $\CC_A$ is anomalous.  On the other hand, under $\CC_S$, it decomposes into two rank-three tensors under the $\CC_S$-invariant subgroup $SO(6)$. Each has the ``self-duality'' constraint
\begin{align}
A_{\pm}^{ijk} = \pm \frac{i}{3!} \epsilon^{ijklmn} A_{\pm}^{lmn} \,.
\end{align}
Note that there is no distinction between upper and lower indices under $SO(6)$.  In an $SO(6)$ instanton background, there are six zero modes for each, and hence $\CC_{S}$ is the symmetry of the theory.

Going back to the original $SU(6)$, the $\CC_{S}$ is given by a symmetric $C_{ij}$ matrix
\begin{align}
A^{ijk} \rightarrow \frac{i}{3!} \epsilon^{ijklmn} C_{lr} C_{ms} C_{nt} A^{rst} \,,
\end{align}
where we chose one possible sign.  Note that a factor of $i$ is needed to ensure $\CC_{S}^{2} = 1$ in the basis where $C_{ij} = \delta_{ij}$.  Then a charge-conjugation invariant $D$-flat direction is
\begin{align}
A^{123} = v \,, \qquad
A^{456} = -i v \,.
\end{align}
Note that the $D$-flatness requires only $D = |A^{123}|^{2} - |A^{456}|^{2} = 0$ and it does not fix the relative phase between the two expectation values.  It is the charge conjugation invariance that fixes the relative phase.\footnote{If the relative phase is different, we choose a different basis to define a new $C_{ij} = \delta_{ij} e^{i\alpha}$ and the charge conjugation invariance always holds.}  It breaks $SU(6)$ to $SU(3)_{1} \times SU(3)_{2}$, and the low-energy gauge coupling constants are given by decoupling the heavy vector multiplet of mass $v$ or $-iv$, and hence
\begin{subequations}
\begin{align}
\frac{8\pi^{2}}{g_{1}^{2}} + i \theta_{1}
&= \frac{8\pi^{2}}{g_{6}^{2}} + i\theta_{6} + 6 \ln \frac{v}{\mu} \,, \\[5pt]
\frac{8\pi^{2}}{g_{2}^{2}} + i \theta_{2}
&= \frac{8\pi^{2}}{g_{6}^{2}} + i\theta_{6} + 6 \ln \frac{-i v}{\mu} \,.
\end{align}
\end{subequations}
As a result, the low-energy couplings differ by $i(\theta_{1}-\theta_{2}) = i \pi$.  Therefore, the two signs in Eq.~\eqref{eq:sign} correspond to different low-energy $SU(3)$ gauge groups.  Namely, it is not an ambiguity; both appear in the low-energy theory.

Each $SU(3)$ factor develops the gaugino condensate
\begin{align}
W = \omega^{n_1} \Lambda_{3,1}^{3} + \omega^{n_2} \Lambda_{3, 2}^{3} = (\omega^{n_1} - \omega^{n_2}) \frac{\Lambda_6^5}{(A^4)^{1/2}} \,.
\end{align}
Here, $\omega = e^{2\pi i/3}$ is the cubic root of unity, and $n_{1,2} = 0, 1, 2$, leading to nine configurations.  Note the relative minus sign that comes from the difference in vacuum angles $\theta_{1}-\theta_{2} =\pi$ in the gauge coupling constants of $SU(3)_{1,2}$.

The charge conjugation is the interchange of two $SU(3)$ factors.  When $n_{1} \neq n_{2}$, the charge conjugation is broken and therefore the configurations are paired $(n_{1}, n_{2}) \leftrightarrow (n_{2},n_{1})$, and also the theory has run-away behavior.  When $n_{1} = n_{2}$, on the other hand, the superpotential cancels exactly and there is a moduli space of vacua.  The charge conjugation is unbroken in this case.

The UV theory has a ${\mathbb Z}_{6}$ discrete symmetry of changing the phase of $A^{ijk}$ by the sixth root of unity because of its Dynkin index six.  However, this symmetry is spontaneously broken.  This is because two gaugino condensates can be parameterized by expectation values of $W_{\alpha}W^{\alpha}$ and ${\cal O} = \epsilon_{ijklmn} A^{ijk} A^{lmr} (W_{\alpha})_{r}^{s} (W^{\alpha})_{s}^{n}$, where the latter breaks ${\mathbb Z}_{6}$ to ${\mathbb Z}_{2}$ \cite{Csaki:1997aw}.  Therefore ${\mathbb Z}_{6}$ anomalies are not matched mod 6, but are only matched mod 2.  The anomalies of charge conjugation are matched correctly (\cref{tab:SU6rank32}).

\section{Conclusion}
\label{sec:conc}

In this paper, we investigated issues of anomalies associated with the outer automorphisms of the gauge groups. We discovered a variety of interesting issues. In some cases, outer automorphisms themselves can be anomalous and hence are not symmetries of the theory. This resolves some paradox about charge-conjugation invariance of chiral symmetry breaking order parameter. When the outer automorphism is non-anomalous, it serves a role in 't Hooft anomaly matching conditions between UV and IR theories, or for duality between electric and magnetic theories. In all cases we studied, they do match perfectly. When the anomaly matching appears to fail, it serves as an indication that parts of the global symmetries are broken. We discussed such cases as well. In all cases, studying anomalies associated with the outer automorphisms are useful and even necessary. This is the main result of this paper.

There is an immediate consequence of our result. The authors of the papers \cite{Bourget:2018ond, Arias-Tamargo:2019jyh, Arias-Tamargo:2021ppf} discussed gauging the {\it principal extension}\/ $SU(N) \rtimes \CC$. Obviously, this is possible only when $\CC$ is non-anomalous under $SU(N)$. For the $\mathcal{N}=2$ SQCD theories in four dimensions that these works focused on, $\CC$ is guaranteed to be non-anomalous. But for a more general construction, this constraint needs to be considered.  Furthermore, we have shown that the technique introduced in \cite{Arias-Tamargo:2019jyh} of finding possible principle extensions of the gauge group $G=SU(N)$ via the Cartan classification of symmetric spaces does not work for the case $G=SO(2k)$.

There are possible future directions concerning outer automorphisms. In topological insulators, the boundary states may be Majorana fermion. The presence of such a state is protected by topology. It would be interesting to see if it has a connection to anomalies under charge conjugation. If the bulk theory is anomalous under charge conjugation, it needs to be accompanied by edge states that cancel the anomaly. It may provide an alternative argument for topologically protected states. See a related discussion concerning time reversal in \cite{Gaiotto:2017yup}.

We restricted the background gauge fields to configurations that are invariant under the outer automorphisms to study well-defined transformation properties of path integral measures. It would be also interesting to see if this restriction can be relaxed. We briefly discussed difficulties associated with such an effort in the appendix.

We believe our paper is only the beginning of studies of outer automorphism anomalies.

\acknowledgments
H.\,M.\ thanks Csaba Cs\'aki for an encouragement to write up this paper. We are grateful to the anonymous referee for helpful comments.
B.\,H.\ is supported by the Swiss National Science Foundation, under grants no. PP00P2-170578 and no. 200020-188671, and through the National Center of Competence in Research SwissMAP.
X.\,L.\ is supported by the U.S. Department of Energy, under grant number DE-SC0011640.
T.\,M.\ is supported by the World Premier International Research Center Initiative (WPI) MEXT, Japan, and by JSPS KAKENHI grants JP19H05810, JP20H01896, and JP20H00153.
The work of H.\,M.\ was supported by the Director, Office of Science, Office of High Energy Physics of the U.S. Department of Energy under the Contract No. DE-AC02-05CH11231, by the NSF grant PHY-1915314, by the JSPS Grant-in-Aid for Scientific Research JP20K03942, MEXT Grant-in-Aid for Transformative Research Areas (A) JP20H05850, JP20A203, by WPI, MEXT, Japan, and Hamamatsu Photonics, K.K.

\appendix
\section{Non-Self-Conjugate Gauge Fields}
\label{ap:ap1}

In this paper, we studied anomalies associated with outer automorphisms by restricting to the gauge field configurations (both for the dynamical gauge field and those associated with weakly gauged global symmetries with spectators) that are self-conjugate (invariant) under the outer automorphism. This is because the question whether
\begin{equation}
\int \mathcal{D}\psi \mathcal{D}\bar{\psi}\, \exp\left[i\int dx\, \bar{\psi}\, i\slashed{D}(A)\, \psi\right]
\stackrel{C}{\longrightarrow}
\pm \int \mathcal{D}\psi \mathcal{D}\bar{\psi}\, \exp\left[i\int dx\, \bar{\psi}\, i\slashed{D}(A)\, \psi\right] \,,
\end{equation}
due to the property of the measure is a well-defined question we could study by working out eigenvalues under outer automorphism for each eigenmode of the Dirac operator. However, one may wonder if stronger constraints can be obtained by considering gauge field configurations that are not self-conjugate configurations.

Obviously for such not self-conjugate gauge fields, we are looking at
\begin{align}
\int \mathcal{D}\psi \mathcal{D}\bar{\psi}\, \exp\left[i\int dx\, \bar{\psi}\, i\slashed{D}(A)\, \psi\right]
\stackrel{C}{\longrightarrow}
\pm \int \mathcal{D}\psi \mathcal{D}\bar{\psi}\, \exp\left[i\int dx\, \bar{\psi}\, i\slashed{D}(A^C)\, \psi\right] \,,
\end{align}
where $A^{C}$ is the charge-conjugated gauge field. The question then is what happens when $A^{C}$ is deformed smoothly back to $A$. If we perfectly understand the spectral flow, namely the continuous change of the eigenvalues of the Dirac operator for $A_{t} = (1-t) A^{C} + t A$, we can probably extend the discussion to non-self-conjugate gauge field configurations. This is beyond the scope of this paper.

\bibliographystyle{JHEP}
\bibliography{ref}

\end{document}